\documentclass[preprints,article,accept,moreauthors,pdftex,10pt,a4paper]{mdpi} 

\firstpage{1} 
\pubvolume{xx}
\issuenum{1}
\articlenumber{5}
\pubyear{2019}
\copyrightyear{2019}
\history{}
\pdfoutput=1

\Title{Gamma-ray sensitivity to dark matter subhalo modelling at high latitudes}

\newcommand{\Jf}{$\mathcal{J}$-factor}
\newcommand{\Fermi}{\textit{Fermi}}

\newcommand{\Msun}{$M_\odot$}
\newcommand{\sv}{\langle \sigma v \rangle}

\Author{Francesca Calore$^{1}$\orcidA{}, Moritz H\"utten $^{2}$\orcidB{} and Martin Stref $^{1,3}$\orcidC{}}

\AuthorNames{Francesca Calore, Moritz H\"utten and Martin Stref}

\address{%
$^{1}$ \quad  Univ.~Grenoble Alpes, USMB, CNRS, LAPTh, F-74000 Annecy, France \\
$^{2}$ \quad  Max-Planck-Institut f\"{u}r Physik, F\"{o}hringer Ring 6, D-80805 M\"{u}nchen, Germany\\
$^{3}$ \quad  Laboratoire Univers \& Particules de Montpellier (LUPM), CNRS \& Université de Montpellier,
Place Eugène Bataillon, F-34095 Montpellier CEDEX 05, France\\
}

\corres{Correspondence: calore@lapth.cnrs.fr}

\abstract{Searches for ``dark" subhaloes in gamma-ray point-like source catalogues are among promising strategies for indirect dark matter detection. 
Such a search is nevertheless affected by uncertainties related, on the one hand, to the modelling of the dark matter subhalo distribution in Milky-Way-like galaxies, and, on the other hand, to the sensitivity of gamma-ray instruments to the dark matter subhalo signals. In the present work, we assess the detectability of dark matter subhaloes in \Fermi-LAT catalogues, taking into accounts uncertainties associated with the modelling of the Galactic subhalo population. We use four different halo models bracketing a large set of uncertainties. For each model, adopting an accurate detection threshold of the LAT to dark matter subhalo signals and comparing model predictions with the number of unassociated point-sources in \Fermi-LAT catalogues, we derive upper limits on the annihilation cross section as a function of dark matter mass. Our results show that, even in the best-case scenario (i.e.~\texttt{DMonly} subhalo model), which does not include tidal disruption from baryons, the limits on the dark matter parameter space are less stringent than current gamma-ray limits from dwarf spheroidal galaxies.
Comparing the results obtained with the different subhalo models, we find that baryonic effects on the subhalo population are significant and lead to dark matter constraints that are less stringent by a factor of $\sim$2 to $\sim$5. This uncertainty comes from the unknown resilience of dark matter subhaloes to tidal disruption. }


\begin{document}

\section{Introduction}
The identification of dark matter (DM) is one of the major endeavours of particle physics and cosmology of the XXI$^{\rm st}$ century.
Despite theoretical and experimental efforts deployed to detect DM particles, the nature of this elusive form of matter remains mostly unknown. We know cold DM to be successful in describing the Universe on large scales~\cite{Aghanim:2018eyx}. However, deviations from cold DM predictions on small scales tantalise this paradigm and cast serious doubts on the weakly interacting massive particle (WIMP) hypothesis, the most scrutinised model for cold DM so far~\cite{Bertone:2004pz}. Additionally, searches for DM particle candidates at the weak scale have been until now unsuccessful with current instruments, on ground and in space. In particular, attempts of indirect detection of high-energy photons from WIMP self-annihilation provide some among the strongest limits on WIMP DM~\cite{Bringmann:2012ez,2016ConPh..57..496G}.
At this stage, it is unclear if the WIMP (and cold DM) paradigm has to be revised in favour of other, still viable, DM particle models (warm and ultra-light DM models), or if it is instead kinematically outside of the main explored range and can be discovered with the next generation of gamma-ray telescopes, e.g. the Cherenkov Telescope Array (CTA, \cite{Acharya:2013sxa}).

Indirect detection constraints on the WIMP parameter space are unavoidably affected by background model systematics. This is particularly severe in the inner region of the Galaxy, where the gamma-ray emission is dominated by the interactions of cosmic rays with the interstellar matter and fields (i.e. Galactic diffuse emission). ``Cleaner" and, in this respect, more promising targets for DM identification are dwarf spheroidal galaxies, optically faint galaxies whose dynamics has been proved to be dominated by large haloes of DM~\cite{Strigari:2018utn}. Those faintest detectable galaxies can probe the WIMP paradigm with multi wavelength observations: from optical to gamma rays (see for example~\cite{Ackermann:2015zua,Calore:2018sdx}). Moreover, the DM haloes hosting dwarf spheroidal galaxies are thought to be the most massive of a vast population of DM subhaloes, overdensities in the DM host halo surrounding our Galaxy \cite{2008MNRAS.391.1685S,2008Natur.454..735D,2019Galax...7...81Z}. While the majority of these DM subhaloes lacks an optical counterpart, a steady gamma-ray signal from directions where no object can be associated in other wavelengths would be a hint for WIMP annihilation.

Searches for DM subhaloes are typically performed in \Fermi-LAT catalogues. Point source catalogues like the 
3FGL \cite{Acero:2015gva} and the 2FHL \cite{Ackermann:2015uya} contain a number of gamma-ray sources which are not associated with any known astrophysical object. Classification algorithms, utilizing in particular spectral information, are applied on these unassociated sources in order to single out potential DM subhalo candidates \cite{Mirabal:2016huj,Salvetti:2017nkp,Coronado-Blazquez:2019puc}.\\

Limits on the DM parameter space (annihilation cross section vs mass) are derived by comparing the number of expected DM subhalo candidates in the catalogues with predictions of the number of subhaloes above the \Fermi-LAT detection threshold expected from theoretical models of subhaloes~\cite{Belikov:2011pu,Berlin:2013dva,Bertoni:2015mla,Schoonenberg:2016aml,Hooper:2016cld,Calore:2016ogv,Coronado-Blazquez:2019puc}. To this end, one needs to know how many subhaloes are expected to be bright enough in gamma rays to be seen above the standard astrophysical background. This requires, on the one hand, a detailed description of the Galactic subhalo population. The complicated physics of subhalo evolution inside the potential of their host leads to different quantitative pictures depending on the models. To bracket these uncertainties, various models, either analytical or based on numerical simulations, are considered in this study, see Sec.~\ref{sec:subhalo_models}.
On the other hand, the number of expected detectable subhaloes is obtained by convolving the DM subhalo signal with the \Fermi-LAT detection threshold. The LAT detection threshold depends on the spectral signal that is looked for. 
Ref.~\cite{Calore:2016ogv} showed that computing the sensitivity of the LAT to DM subhalo signals, adopting the specific spectral energy distribution determined by the particle physics DM model (see also Sec.~\ref{sec:gamma_rays_from_subhalos}), 
provides more accurate predictions on the number of expected detectable subhalo and that important differences with respect to assuming a fixed sensitivity threshold arise. We will therefore use the \Fermi-LAT detection threshold as derived in~\cite{Calore:2016ogv}.

The goal of the present paper is to assess the detectability of DM subhaloes as predicted by state-of-the-art DM subhaloes models~\cite{Hutten:2019tew}.
We will do so by using the more accurate \Fermi-LAT sensitivity threshold to DM subhalo signals~\cite{Calore:2016ogv}.
In Sec.~\ref{sec:subhalo_models} we describe the Galactic subhalo models, in Sec.~\ref{sec:gamma_rays_from_subhalos} we remind the reader the main ingredients to compute the gamma-ray DM signal from dark subhaloes, and in Sec.~\ref{sec:sensitivity} how the LAT sensitivity is computed.  We present the results in Sec.~\ref{sec:results}, and conclude in Sec.~\ref{sec:conclusions}.

\section{Galactic subhalo modelling}
\label{sec:subhalo_models}

Subhaloes are subject to a variety of phenomena, including tidal stripping, gravitational shocking and dynamical friction, which make their modelling challenging.
Subhaloes can be studied by the means of fully-numerical cosmological simulations or simplified analytical models.
These different approaches lead to similar qualitative pictures regarding the Galactic subhalo population but often differ on a quantitative level.
To get a handle on the modelling uncertainties, several models are considered in this study. These models share some common features: spherical symmetry is assumed for the Galactic halo, subhaloes all have a Navarro-Frenk-White (NFW) density profile and their mass function is a power law with index $\alpha_{\rm m}=1.9$.\footnote{The mass function is sharply cut at $m_{\rm min}=10^{-6}\,M_{\odot}$. This mass cutoff can be related to the kinetic decoupling of the DM particle in the early Universe, see e.g.~\cite{Green:2005fa,Bringmann:2009vf}.} 
These assumptions are all verified on the scales resolved by numerical simulations, see e.g.~\cite{Diemand:2008in,Springel:2008b}. 
Four configurations are considered, which are identical to those used in \citet{Hutten:2019tew}, which the reader is referred to for further details.

Our first model is based on the \textit{Aquarius} DM-only N-body simulation \cite{Springel:2008b} and as such is called \texttt{DMonly}. The subhalo spatial distribution in \textit{Aquarius} is found to be well fitted by an Einasto profile with parameters $\alpha_{\rm E}=0.68$ and $r_{-2}=199$ kpc. The core in the distribution is created by tidal interactions which tend to disrupt subhaloes at the center of the host.
The total number of clumps is set to 300 above a mass of $10^{8}\,\rm M_{\odot}$ 
Subhaloes are further assumed to follow the mass-concentration relation given by \citet{Moline:2016pbm}.
While a well-known effect of tides is to remove matter from the outskirts of subhaloes, this is not accounted for in \texttt{DMonly} and all the subhaloes have their cosmological extension (defined with respect to the critical density of the Universe).

The \texttt{Phat-ELVIS} model is based on a suite of DM-only simulations which incorporate a static disc potential~\cite{Kelley:2018pdy}. Through gravitational shocking, the disc is very efficient at disrupting most subhaloes in the inner 30 kpc of the host galaxy. The spatial distribution of the remaining population is well fitted by the following function:
\begin{eqnarray}
\frac{\mathrm{d}P}{\mathrm{d}V}(r) = \frac{A}{1+e^{-(r-r_{0})/r_{\rm c}}}
\times\exp\left\{-\frac{2}{\alpha}\left[\left(\frac{r}{r_{-2}}\right)^{\alpha}-1\right]\right\}\,,
\end{eqnarray}
with $\alpha=0.68$, $r_{0}=29.2$ kpc, $r_{\rm c}=4.24$ kpc and $r_{-2}=128$ kpc. Similar to the \texttt{DMonly} model, the mass-concentration relation is taken from \cite{Moline:2016pbm} and the density profile of subhaloes extends to their cosmological extension.

Our next configurations are based on the semi-analytical model of \citet{Stref:2016uzb} (SL17 from now on). This model relies on a realistic description of the Milky Way and incorporates the stripping effect due to the gravitational potential of the Galaxy as well as the shocking effect from the disc.
It is not clear yet whether the efficient disruption of DM subhaloes as observed in simulations is realistic or not \cite{vandenBosch:2017ynq,vandenBosch:2018tyt,2019arXiv190601642E}.
This can be of importance because it has been shown to impact predictions for indirect searches \cite{galaxies7020065}. 
To account for this uncertainty, we consider two scenarios. In the first one, called \texttt{SL17-fragile}, subhaloes are disrupted when their tidal radius $r_{\rm t}$ is equal to their scale radius $r_{\rm s}$.
In the second one, called \texttt{SL17-resilient}, subhaloes are more robust and survive unless $r_{\rm t}<10^{-2}\,r_{\rm s}$.
Unlike the \texttt{DMonly} and \texttt{Phat-ELVIS} models, subhaloes in the SL17 configurations are stripped down to their tidal radius.

\medskip

Knowing the DM subhalo spatial density $\rho_{\rm DM}$, it is possible to compute the so-called astrophysical or \Jf~towards the direction --line of sight (l.o.s.)-- of the subhalo of interest: 
\begin{equation}
\int_{0}^{\Delta \Omega}\!\!\int_{\rm{l.o.s.}} d\ell \, d\Omega \, \rho_{\rm DM}^2(\ell) \, ,
\label{eq:Jf}
\end{equation}
where the integral along the l.o.s. is further integrated over the solid angle $\Delta \Omega=2\pi\,(1-\cos\,\theta_{\rm int})$. In what follows, we will set $\theta_{\rm int} = 0.1^\circ$ effectively considering subhaloes as point-like sources. We note that previous works have overestimated the \Jf -- and thus got too stringent limits on DM -- by integrating up to 0.5$^\circ$~\cite{Calore:2016ogv} or, up to the DM profile scale radius, e.g.~\cite{Coronado-Blazquez:2019puc}. Indeed, as we will explain below, the way in which the LAT sensitivity to DM spectra is computed strictly applies to point-like sources having an angular extension of 0.1-0.3$^\circ$. Cutting the integration radius up-to 0.1$^\circ$ worsens the final limits on the DM annihilation cross section by a factor of 2, over all the DM mass range. Again, we believe this choice to be truly conservative.
The \Jf~is one of the crucial ingredients to compute gamma-ray DM fluxes, as we will see below.

Having incorporated these models in the CLUMPY code \cite{2012CoPhC.183..656C,Bonnivard:2015pia,Hutten:2018aix}, we consider 1000 Monte Carlo realisations for each configuration, and we select all subhaloes with $\mathcal{J}(<0.1^\circ) > 10^{17} \, \rm GeV^2 cm^{-5}$. The choice of this cut originates from the specific scope of this paper: To show how many subhaloes would be detectable in the \Fermi-LAT catalogues, and what is the role, if any, of low-mass subhaloes.
We note that relying on the simulations done in \cite{Hutten:2019tew} guarantees that the subhalo population is complete in brightness.

\begin{figure}[!h]
	\centering
	\includegraphics[width=0.45\columnwidth]{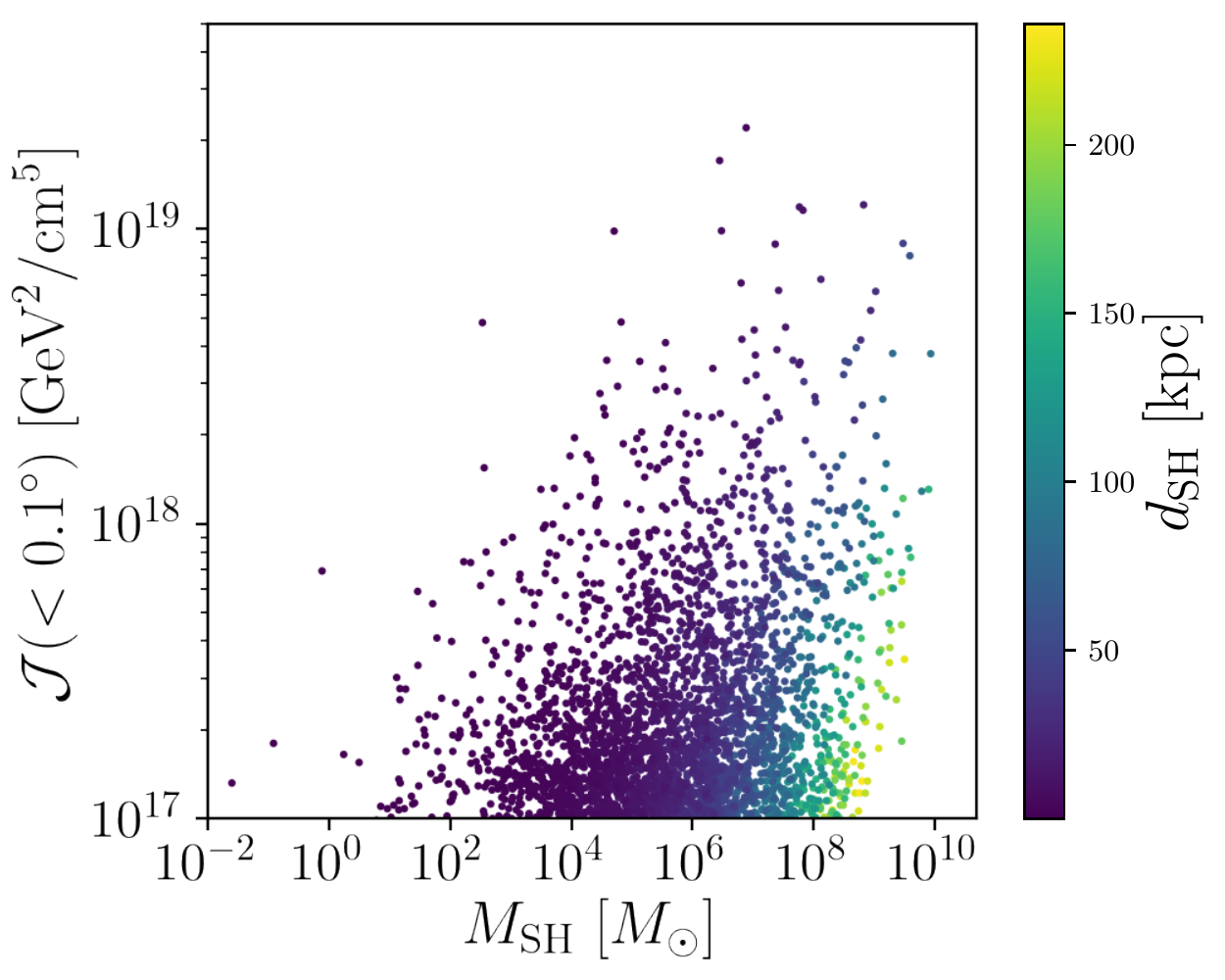}
	\includegraphics[width=0.45\columnwidth]{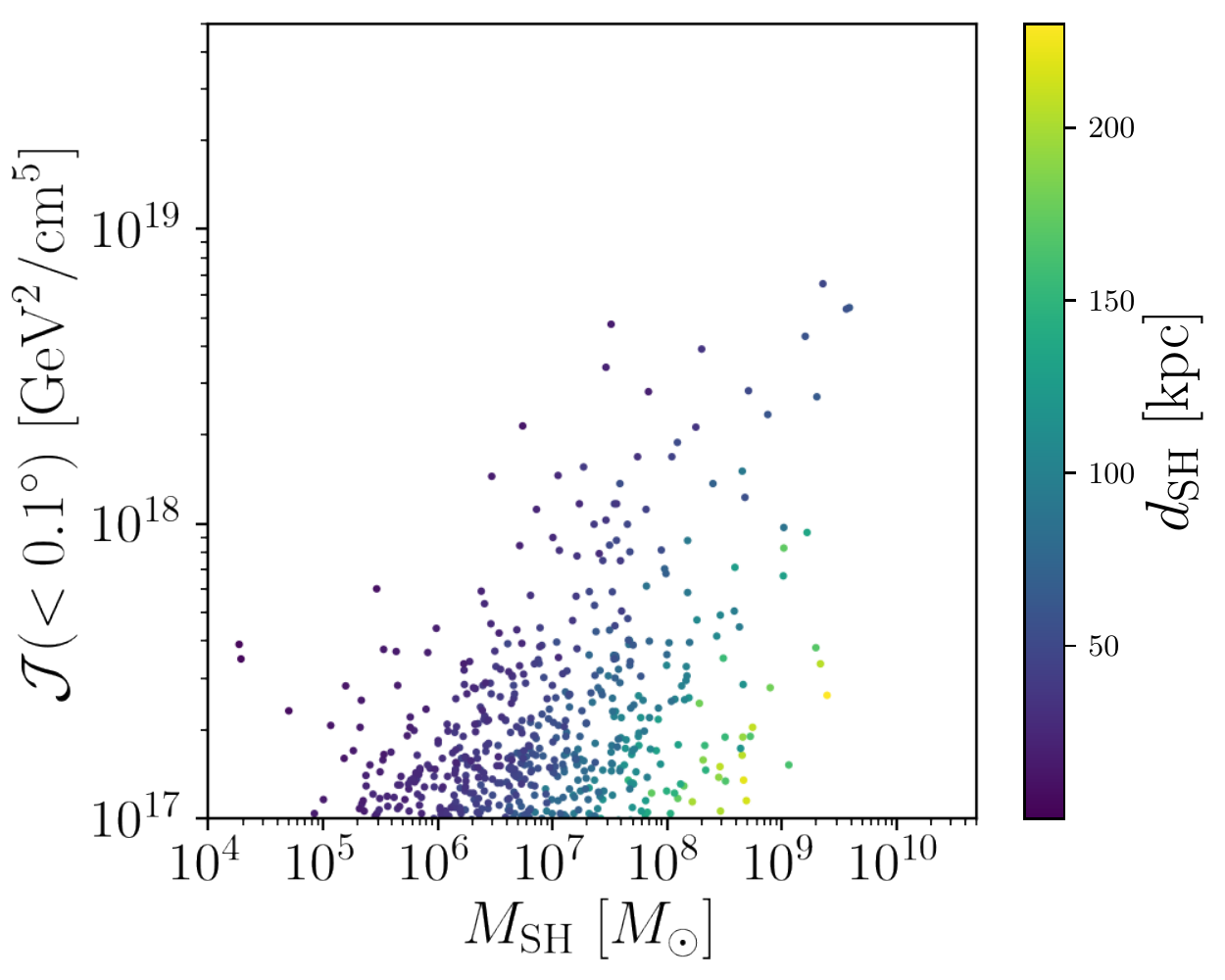}
	\includegraphics[width=0.45\columnwidth]{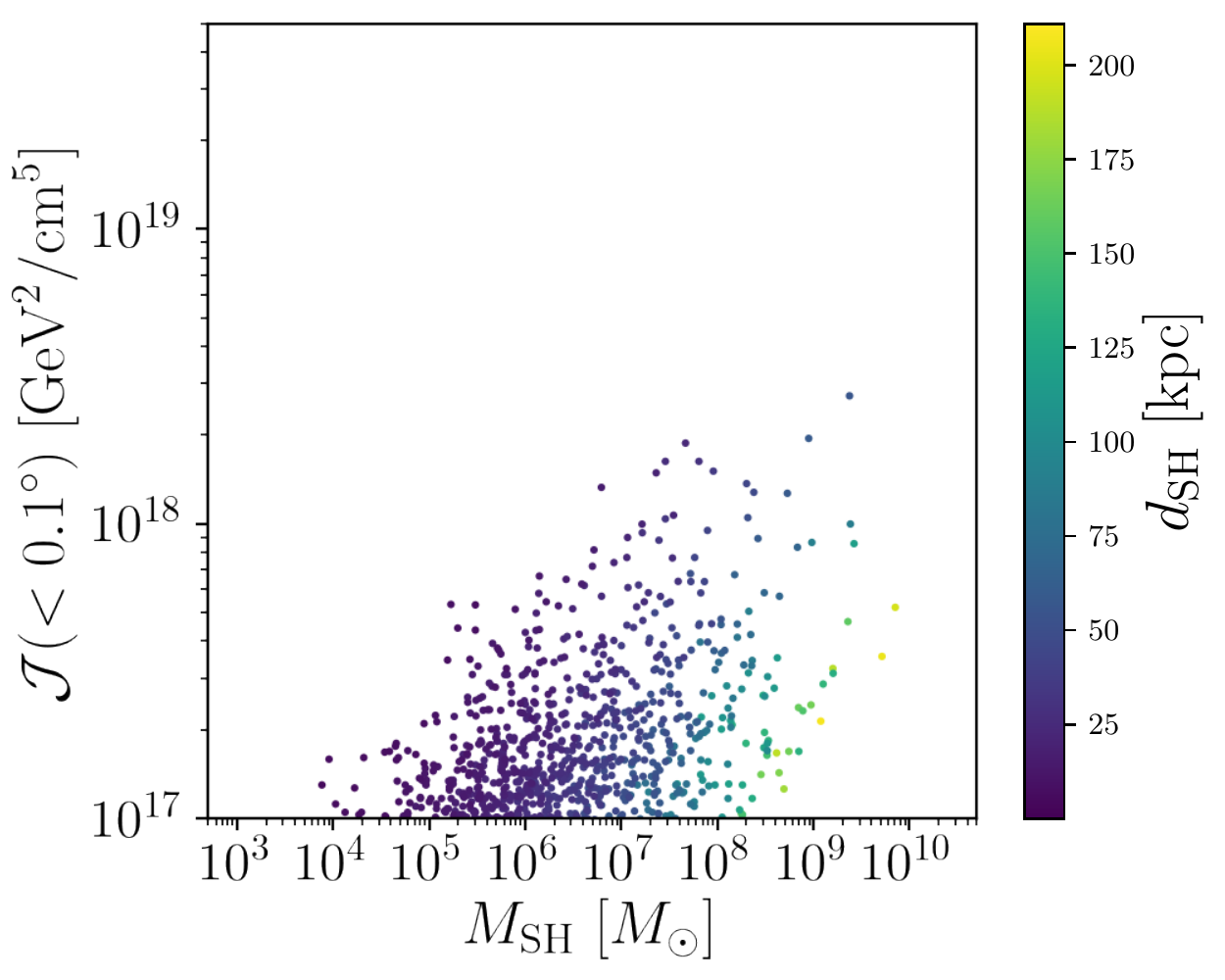}
	\includegraphics[width=0.45\columnwidth]{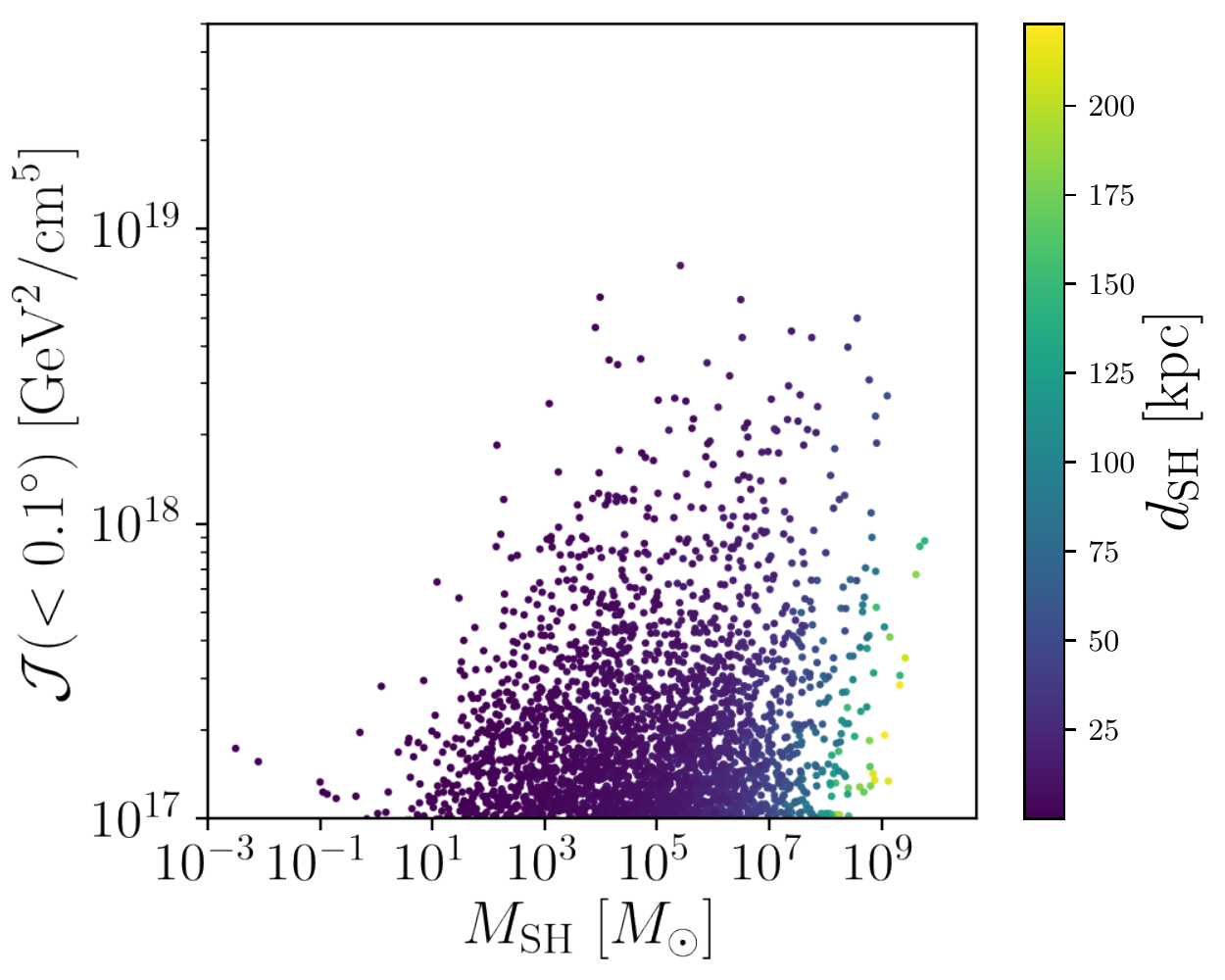}
\caption{Scatter plot of \Jf~values, $\mathcal{J}$ within 0.1$^\circ$, 
	as a function of subhalo mass, $M_{\rm SH}$, in one realisation of the Monte Carlo simulations for each subhalo model --
	top left: \texttt{DMonly}, top right: \texttt{Phat-ELVIS}, bottom left: \texttt{SL17-fragile},  bottom right: \texttt{SL17-resilient}. 
	The colour-bar represents the subhalo distance from Earth, hereafter $d_{\rm SH}$. The realisation shown is the one containing the lowest mass subhalo. 
	We remind that we have applied a cut of $\mathcal{J}(<0.1^\circ) > 10^{17} \, \rm GeV^2 cm^{-5}$.}
\label{fig:JvsM}
\end{figure}

In Fig.~\ref{fig:JvsM}, we show the scatter plots of \Jf~values, $\mathcal{J}(<0.1^\circ)$, as a function of subhalo mass, $M_{\rm SH}$, in one realisation of the Monte Carlo simulations for each subhalo model. 

\section{Gamma rays from subhalos}
\label{sec:gamma_rays_from_subhalos}

The \Jf~is proportional to the predicted gamma-ray flux from WIMP DM annihilation.
We therefore expect that the most-likely detectable subhaloes will be also the ones with the highest \Jf.
However, the sensitivity of a gamma-ray telescope to a DM (or astrophysical) signal does also depend on the gamma-ray spectrum that is looked for -- in general harder spectra (e.g.~blazars) are detected more easily --, as we will see below.
 
To compute the predicted flux from DM annihilation, we have to specify the particle physics content of the underlying DM particle model we consider. In what follows, we provide equations for Majorana DM candidates (such as the neutralino in Supersymmetric extensions of the Standard Model) -- predictions for Dirac DM particles can be obtained by multiplying the flux for an additional factor of 1/2.

The flux of photons expected in a given energy range from annihilation of DM particles of mass $m_{\rm DM}$, distributed spatially following the DM distribution $\rho_{\rm DM}$, writes generally as:
\begin{equation}
\mathcal{F}(E_{\rm min},E_{\rm max}) = \frac{ \sv }{8\pi m^2_{\rm{DM}}} \;  \mathcal{J}\;  \int_{E_{\rm min}}^{E_{\rm max}} \frac{\mathrm{d}N^i_{\rm{DM}}}{\mathrm{d}E} \, dE \, ,
\label{eq:DMflux}
\end{equation}
where $\sv$ is the thermally averaged annihilation cross section, and $\mathrm{d}N^i_{\rm{ DM}}/\mathrm{d}E$ is the energy spectrum providing the number of gamma rays per annihilation of DM in a given final state $i$ (e.g.~$b\bar{b}$, $\tau^+\tau^-$, etc.).
We use tabulated DM spectra from~\cite{Cirelli:2010xx}.

\section{\Fermi-LAT sensitivity to DM subhalos}
\label{sec:sensitivity}
We adopt the flux sensitivity calculation of \citet{Calore:2016ogv}, where the authors provided an accurate calculation of the LAT sensitivity to DM annihilation signals from subhaloes and showed that such a determination of the detection threshold leads to significant differences with respect to adopting a fixed flux threshold.
The \Fermi-LAT source detection simulation of DM subhaloes was performed for the third \Fermi-LAT catalog of point sources (3FGL)~\cite{Acero:2015gva}, and the second catalog of hard \Fermi-LAT sources (2FHL)~\cite{Ackermann:2015uya}. In both catalogues, unassociated sources represent a significant fraction of all detected sources: About 15\% in the 2FHL and 30\% in the 3FGL. 

Interestingly, some gamma-ray emitting DM subhaloes can hide among unassociated sources in the \Fermi-LAT catalogues. In the present work, we assess what is the sensitivity to DM subhalo modelling of the \Fermi-LAT 3FGL and 2FHL catalogues.
 
As can be seen from Figs.~5--8 in~\cite{Calore:2016ogv}, the flux sensitivity threshold of \Fermi-LAT for the 3FGL and 2FHL set-ups depends both on latitude and mass of the DM candidate: Regardless of the annihilation channel, the flux sensitivity threshold decreases by a factor of about 2 between $20^\circ$ and $80^\circ$ in latitude for all masses, for both the 3FGL and 2FHL set-up.
Also, higher (lower) DM masses are more easily detected in the 3FGL (2FHL) set-up, as thoroughly explained in~\cite{Calore:2016ogv}. We note that our sensitivity threshold for the $b\overline{b}$ and $\tau^{+}\tau^{-}$ channels is very similar to the one more recently derived by \citet{Coronado-Blazquez:2019puc}. Given the contamination of the Galactic diffuse foreground, the sensitivity calculation of~\cite{Calore:2016ogv} is truly accurate for $|b| > 20^\circ$. We will therefore consider only subhlaoes at high latitudes. 

 
\section{Results}
\label{sec:results}
To derive the number of detectable subhaloes for a given mass and final state annihilation channel, we compute the corresponding gamma-ray flux (Eq.~\ref{eq:DMflux}) in the same energy range of the catalogue we want to compare with -- $E > 0.1$ GeV for the 3FGL and $E > 50$ GeV for the 2FHL. For each subhalo in the Monte Carlo simulations, we then compare the  DM gamma-ray flux with the \Fermi-LAT sensitivity threshold at the position of the subhalo: A subhalo is detected if its gamma-ray flux is larger than the flux threshold at its position.

In general, the number of detectable subhaloes is almost linearly proportional to the annihilation cross section. 
As found in~\cite{Calore:2016ogv}, the number of detectable subhaloes does not strongly depend on the DM mass for annihilation into bottom quarks, while, because of the harder spectrum, the DM mass is more relevant in the case of annihilation into $\tau$ leptons.
In Tab.~\ref{tab:Ndet_sv}, for each model and catalogue configuration, we provide the annihilation cross section required to have at least one subhalo detectable for annihilation into $b$-quarks and $\tau$ leptons. 
We note that the minimal cross section needed to detect at least one subhalo is about a few $10^{-25}$ for annihilation into $b\bar{b}$ in the 3FGL, in the case of the \texttt{DMonly} model. 
The minimal cross section for \texttt{SL17-resilient} is found to be a factor of $\sim 2$ higher, while it is $\sim 4-5$ higher for the \texttt{PhatELVIS} and \texttt{SL17-fragile} models.
The hierarchy between the models is similar for the 2FHL catalogue and for annihilation into $\tau^+ \tau^-$.
These minimal cross sections exceed current bounds from \Fermi-LAT observations towards dwarf spheroidal galaxies, see e.g~\cite{Fermi-LAT:2016uux}. Dwarf spheroidal galaxies are traditionally believed to give the strongest and most robust limits of the DM parameter space -- although several, independent, works addressed the robustness of such a bound showing that it is prone to uncertainties of a factor of a few mainly because on the uncertainty in the modelling of the foreground at the dwarf position~\cite{Calore:2018sdx} and of the dwarf DM distribution~\cite{Bonnivard:2014kza,Klop:2016lug,Ullio:2016kvy}.

\begin{table}
\centering
\begin{tabular}{ |p{2.5cm}||p{2.2cm}|p{2.2cm}||p{2.2cm}|p{2.2cm}| }
 \hline
 \multicolumn{5}{|c|}{One detectable subhalo cross section [$\rm cm^3/s$]} \\
 \hline
  & 3FGL, $b\bar{b} $& 3FGL, $\tau^+ \tau^-$ & 2FHL, $b\bar{b} $& 2FHL, $\tau^+ \tau^-$ \\
 \hline
\texttt{DMonly}  &  8.80 $\times 10^{-25}$ & 17.25 $\times 10^{-25}$ & 3.81 $\times 10^{-23}$    &  10.52 $\times 10^{-23}$ \\
\texttt{PhatELVIS}   & 34.64 $\times 10^{-25}$ & 76.96 $\times 10^{-25}$   &  18.99 $\times 10^{-23}$ & 50.91 $\times 10^{-23}$\\
\texttt{SL17-fragile}  & 44.50  $\times 10^{-25}$&   100.02 $\times 10^{-25}$ & 28.91  $\times 10^{-23}$& 63.82 $\times 10^{-23}$\\
 \texttt{SL17-resilient} &  19.32  $\times 10^{-25}$&   34.23 $\times 10^{-25}$  & 9.70 $\times 10^{-23}$& 19.30 $\times 10^{-23}$\\
 \hline
\end{tabular}
\caption{Cross section required to have at least one subhalo detectable in the 3FGL (2FHL) catalogue set-up for a 100 GeV (1.5 TeV) DM particle mass.}
\label{tab:Ndet_sv}
\end{table}

\medskip

\begin{figure*}[!h]
	\centering  
	\includegraphics[width=0.45\columnwidth]{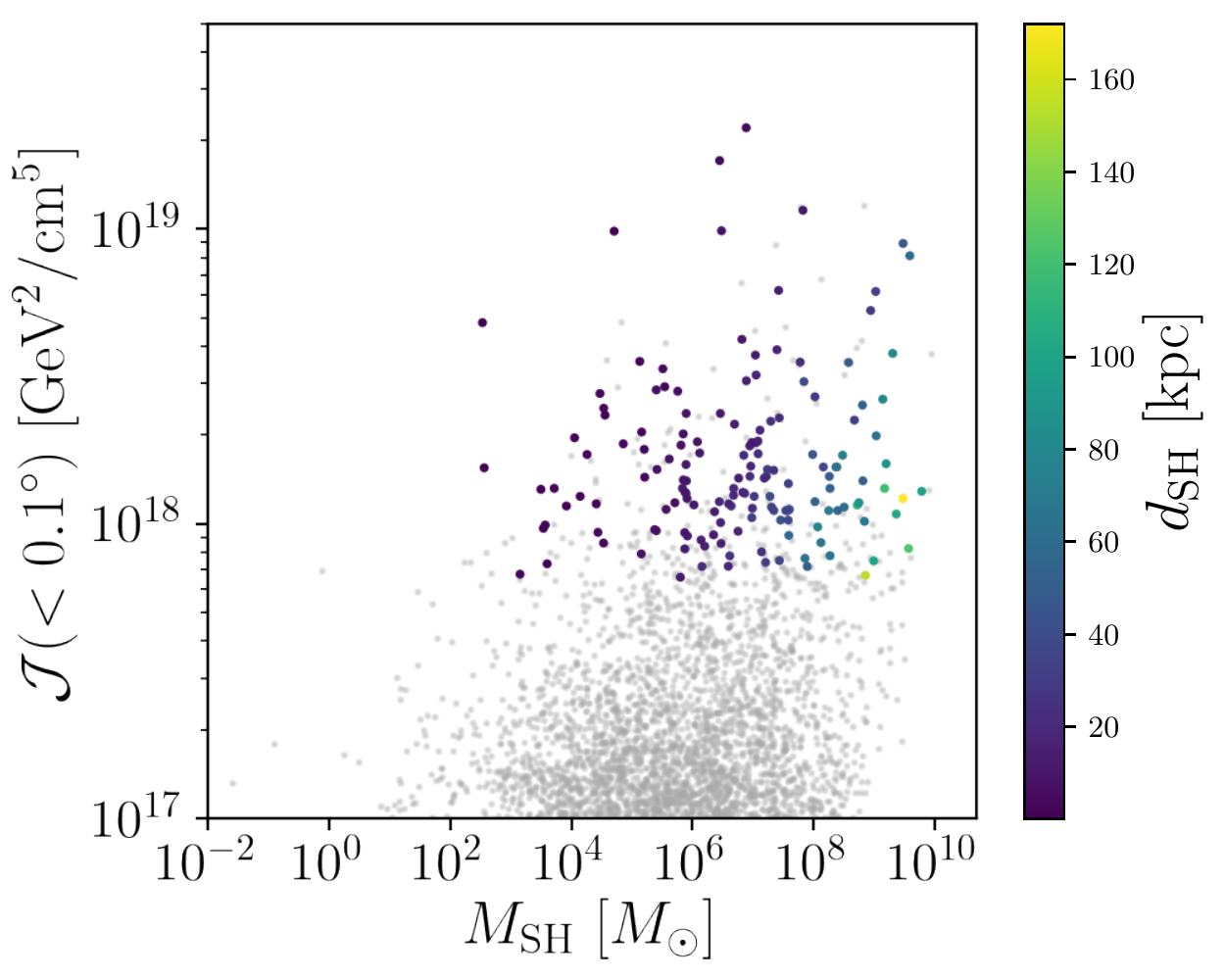}
	\includegraphics[width=0.45\columnwidth]{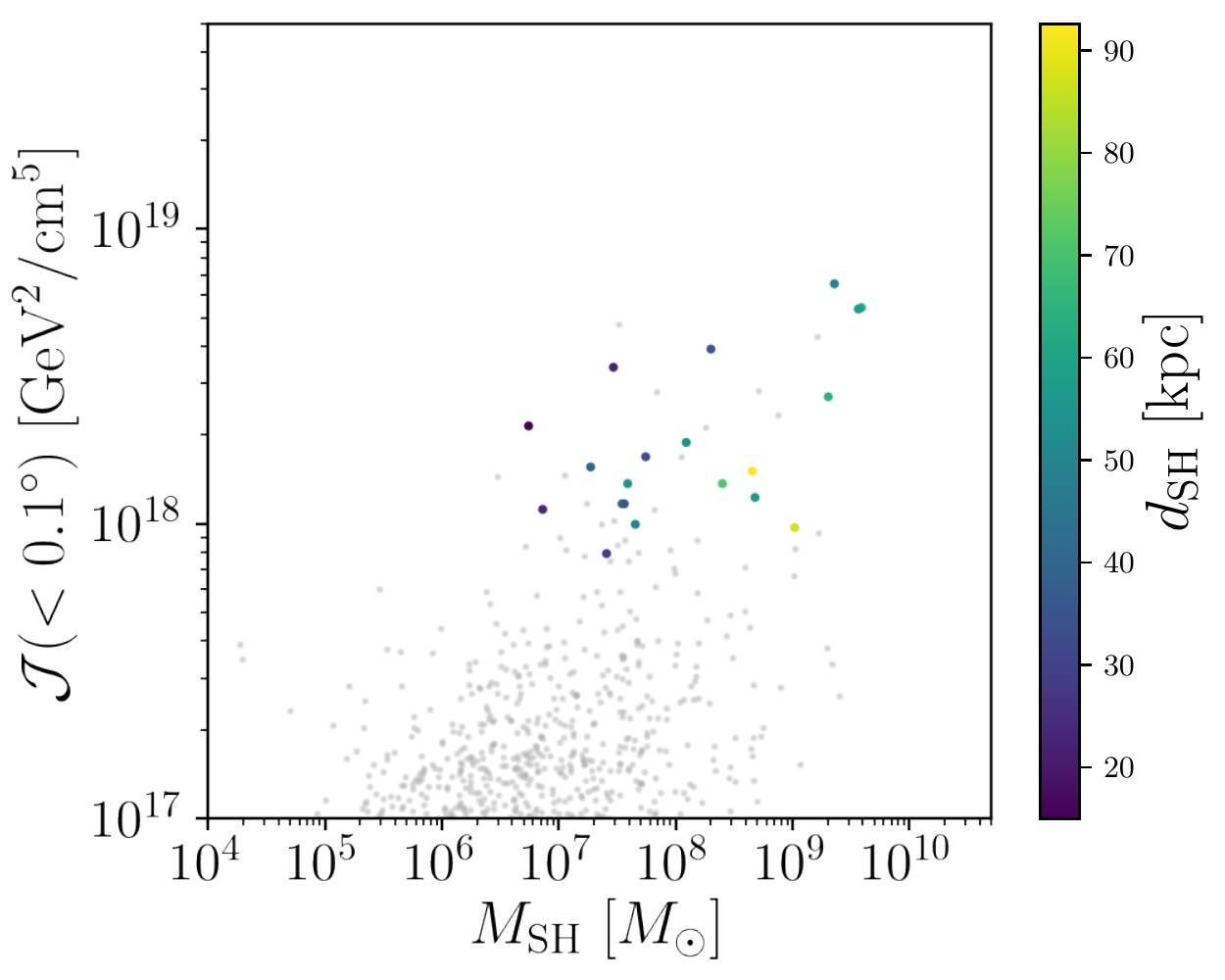}
	\includegraphics[width=0.45\columnwidth]{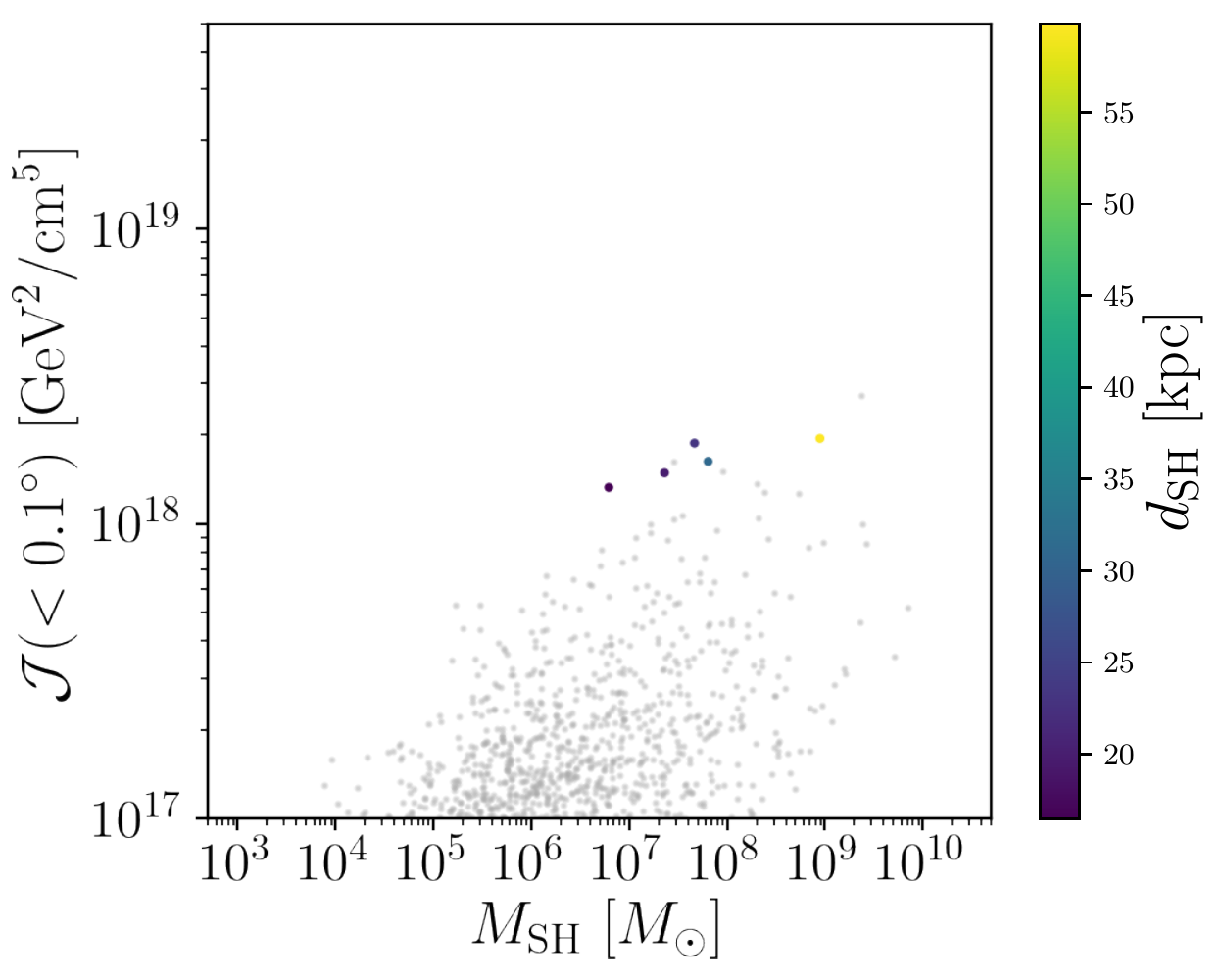}
	\includegraphics[width=0.45\columnwidth]{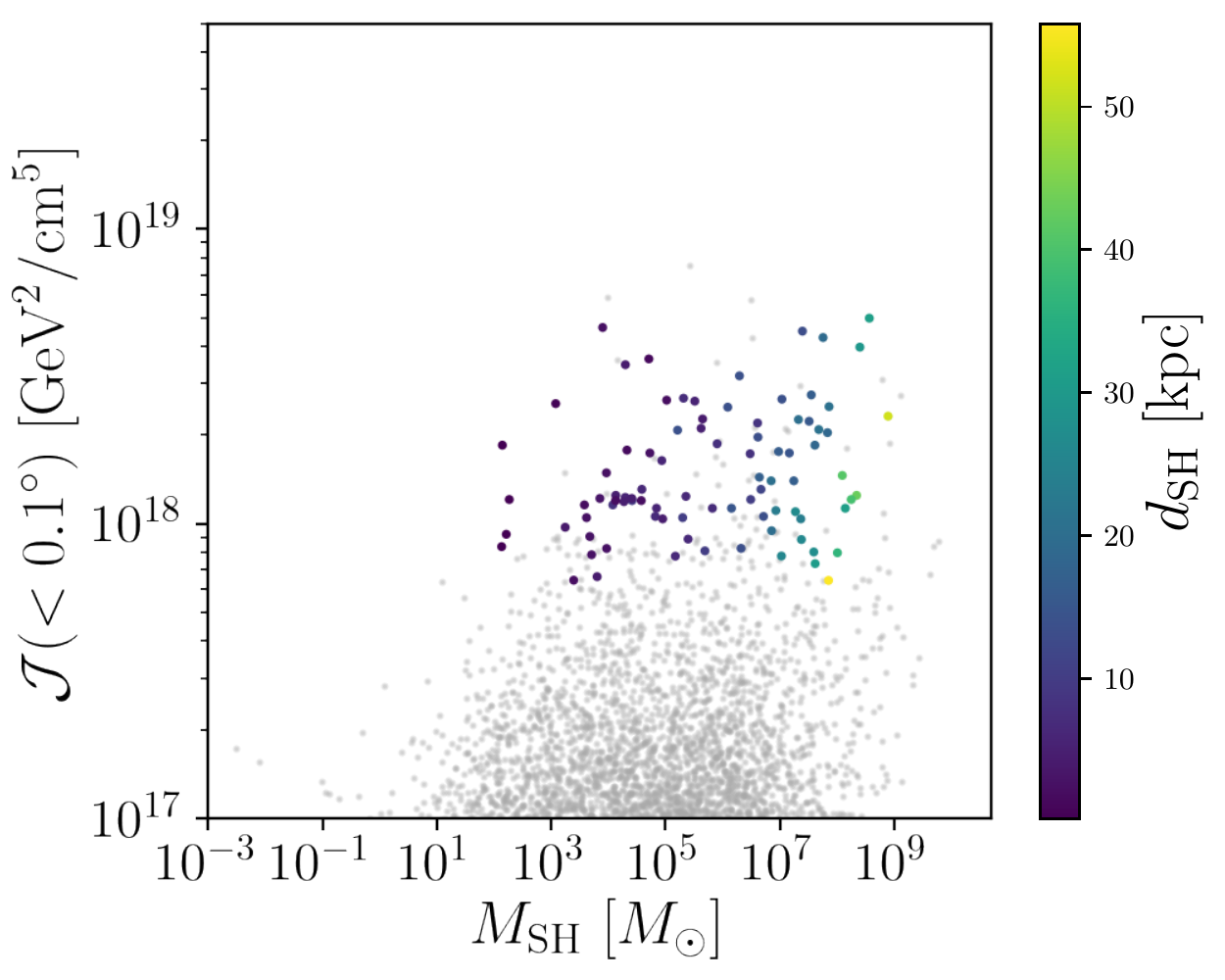}
	\caption{Same as in Fig.~\ref{fig:JvsM} displaying all subhaloes selected as grey dots and those which would be detectable in the 2FHL catalogue as coloured points.
	The results are shown for a DM mass of 100 GeV, $b$-quark annihilation, and a cross section of $5 \times 10^{-24}$ cm$^{3}$/s.
	}
\label{fig:JvsM_detected} 
\end{figure*}

\begin{figure*}[!h]
	\centering  
	\includegraphics[width=0.45\columnwidth]{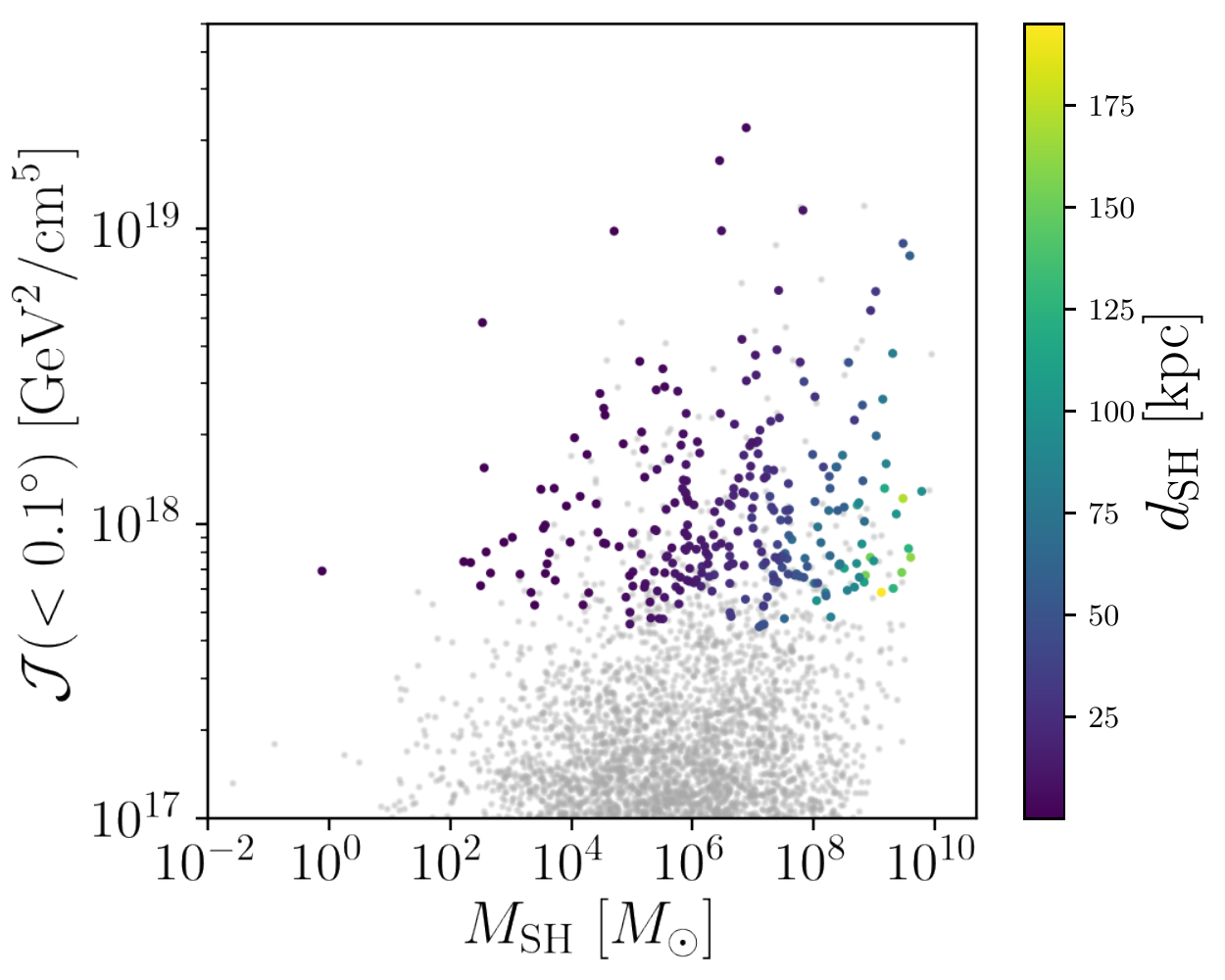}
	\includegraphics[width=0.45\columnwidth]{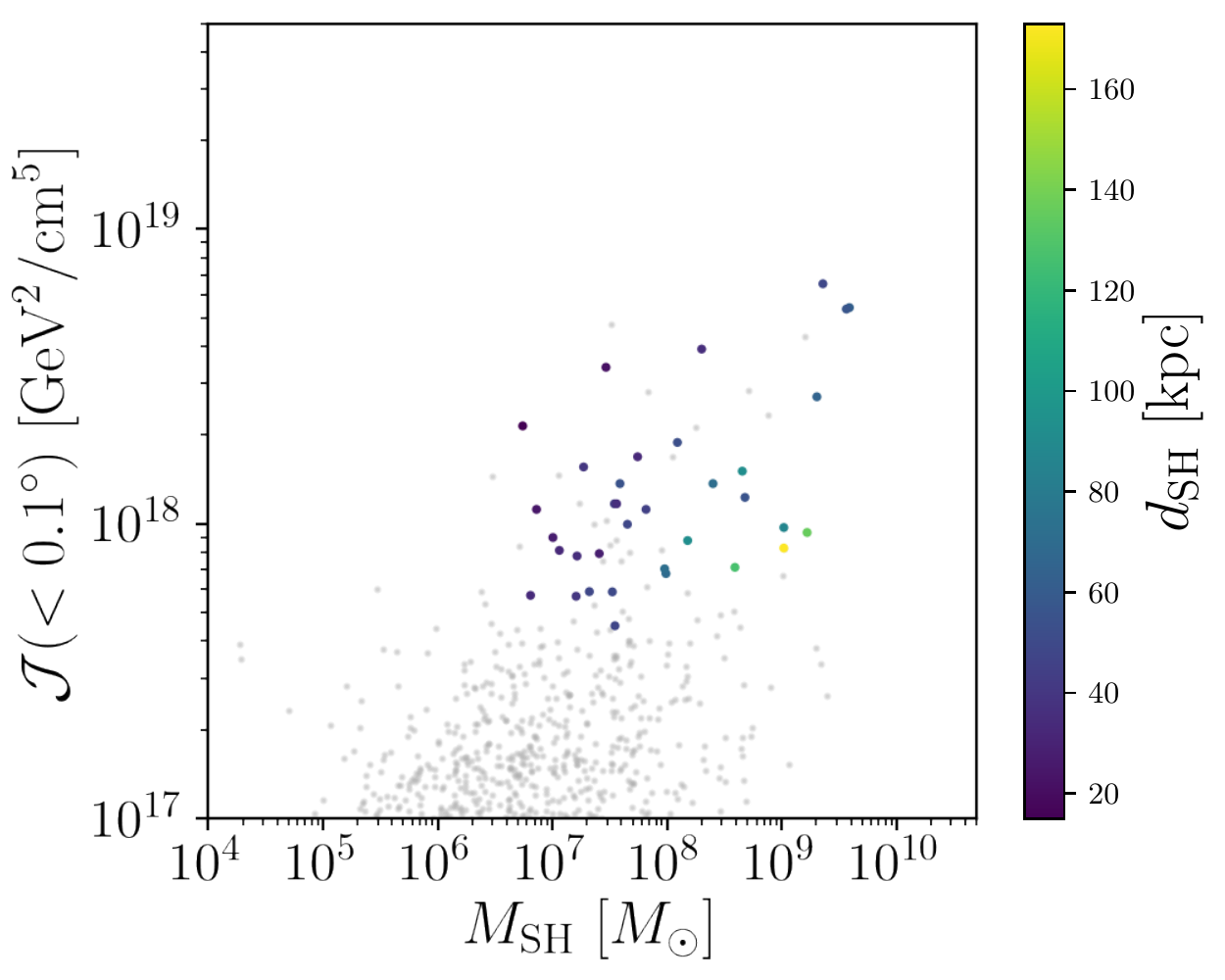}
	\includegraphics[width=0.45\columnwidth]{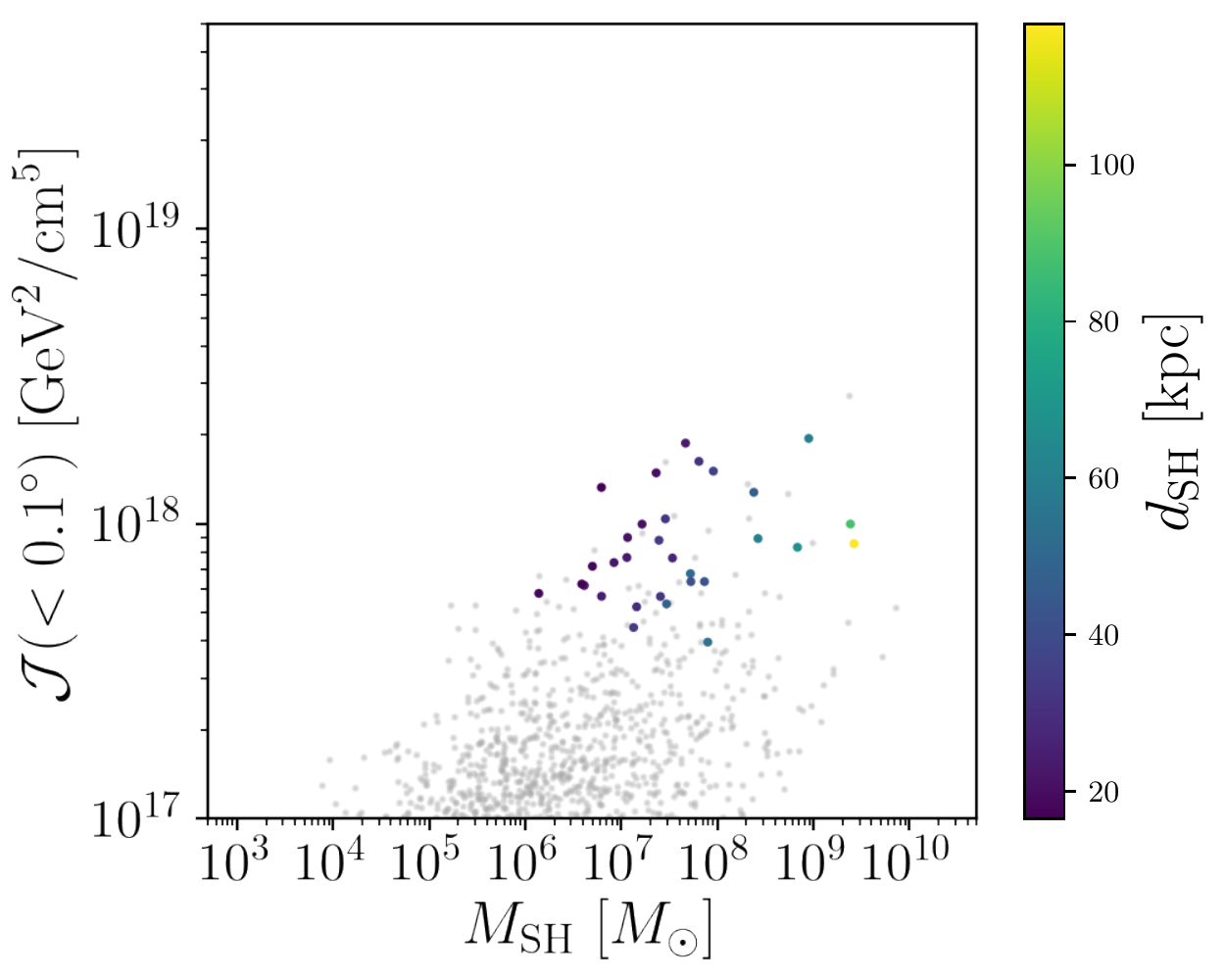}
	\includegraphics[width=0.45\columnwidth]{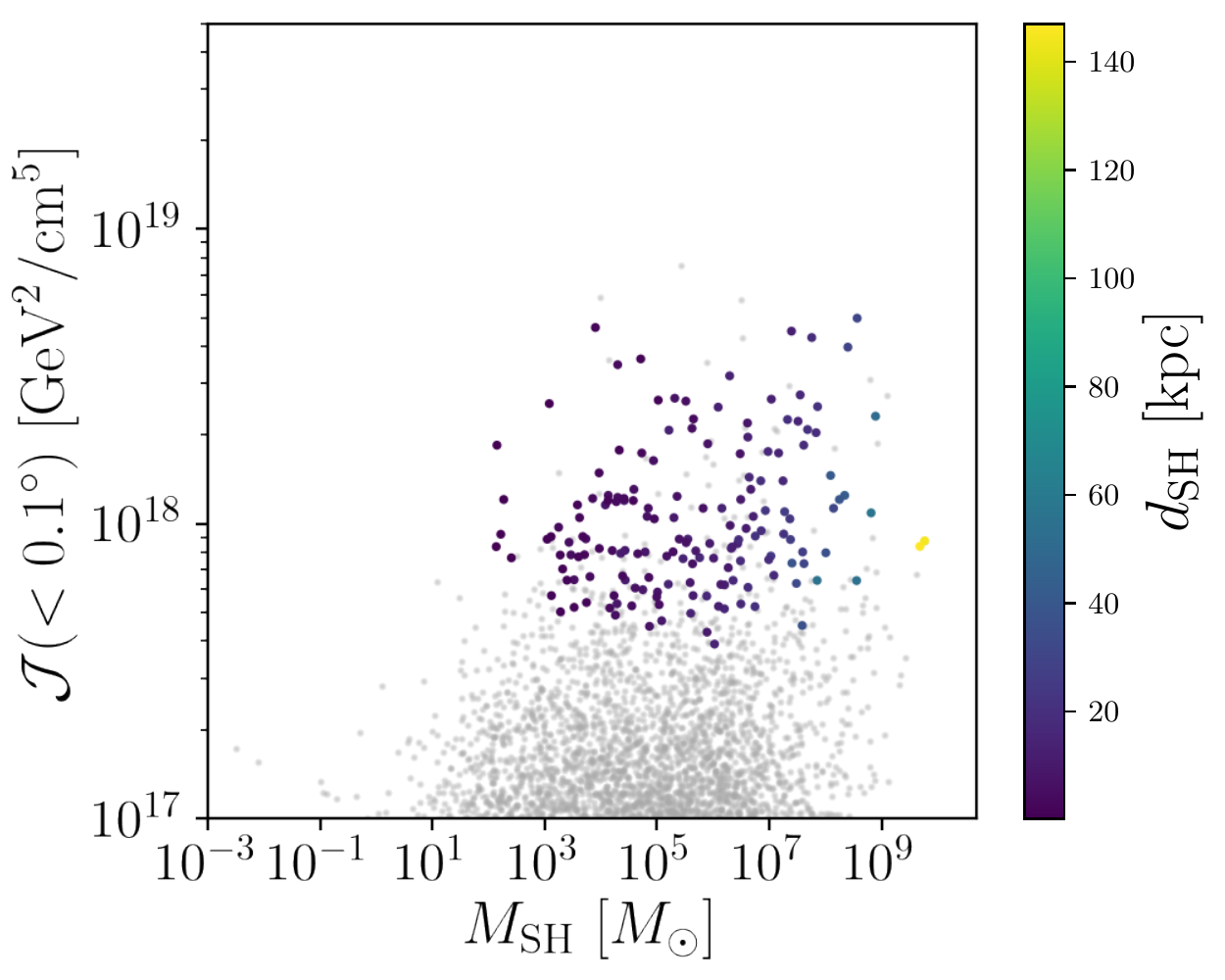}
	\caption{Same as in Fig.~\ref{fig:JvsM} displaying all subhaloes selected as grey dots and those which would be detectable in the 2FHL catalogue as coloured points.
	The results are shown for a DM mass of 1.5 TeV, $b$-quark annihilation, and a cross section of $5 \times 10^{-22}$ cm$^{3}$/s.
	}
\label{fig:JvsM_detected_2FHL} 
\end{figure*}

It is of interest to have a look at the distribution of the \Jf~of detectable subhaloes versus their mass. 
This is shown in Fig.~\ref{fig:JvsM_detected} for the 3FGL catalogue set-up.
In contrast to Fig.~\ref{fig:JvsM}, all subhaloes are here represented by grey dots, while the ones detectable in the 3FGL catalogue are shown by coloured points. Note that these subhaloes, represented by their \Jf s, are detectable for fluxes from a DM particle with mass of 100 GeV and annihilation cross section into $b\bar{b}$  of $5\times 10^{-24}$ cm$^{3}$/s. Fig.~\ref{fig:JvsM_detected_2FHL} shows the same for the 2FHL catalogue set-up, DM mass of 1.5 TeV and annihilation cross section into $b\bar{b}$ of $5\times 10^{-22}$ cm$^{3}$/s
A few considerations are in order. 
First, for the set of particle physics parameter chosen, for the 3FGL (2FHL) set-up the \Jf~threshold for subhalo detection is about $7-8 \times 10^{17} \, (4-5 \times 10^{17}) \rm GeV^2/cm^{5}$ for all four models. 
However, not all subhaloes with \Jf~above this threshold are detectable in the \Fermi-LAT catalogues: Indeed, the DM mass and latitude dependence of the flux sensitivity threshold implies that the highest \Jf~subhaloes sometimes are not the most likely detectable ones with the LAT. This can be clearly seen, for example, in the bottom right panel of Fig.~\ref{fig:JvsM_detected}: While the brightest gamma-ray subhalo has $\mathcal{J} \sim 8 \times 10^{18} \, \rm GeV^2/cm^{5}$, the detectable subhalo with the highest \Jf~has $\mathcal{J} \sim 5 \times 10^{18} \, \rm GeV^2/cm^{5}$. The same occurs for the 2FHL set-up.
Secondly, the mass of detectable subhaloes can span up to seven orders in magnitude (from $\sim 10^2$ M$_{\odot}$ to $\sim 10^{10}$ M$_{\odot}$) depending on the configuration, as it is the case for \texttt{DMonly} and \texttt{SL17-resilient}. We can conclude that among detectable sources there are both dwarf galaxies and dark subhaloes. Indeed, there is a minimum subhalo mass to form a galaxy, which is about $10^{7-8}$ \Msun. Although the exact threshold for star formation is quite debated and dark subhaloes can even coexist with luminous ones above that star formation threshold~\cite{Zhu:2015jwa}, low-mass subhaloes (below  $10^{7}$ \Msun) are almost surely optically dark objects. However, those can still have large \Jf~and be among detectable subhaloes. This occurs for our \texttt{DMonly}, but also in a model where the effect of baryons in the Galaxy is fully modelled (\texttt{SL17-resilient}). 
Finally, we note that in subhalo models where tidal disruption is less efficient (\texttt{DMonly} and \texttt{SL17-resilient}), most detectable subhaloes are located at a distance less than 20 kpc from us. Some of them are even closer than 10 kpc. On the other hand, when tidal disruption is efficient (as in \texttt{PhatELVIS} and \texttt{SL17-fragile}), a larger fraction of detectable subhaloes is located farther away (see also \cite{Hutten:2019tew} for details). This is because the stellar disc disrupts most objects orbiting within the inner $\sim$20 kpc of the galaxy. 

\begin{figure}[!h]
	\centering
	\includegraphics[width=\columnwidth]{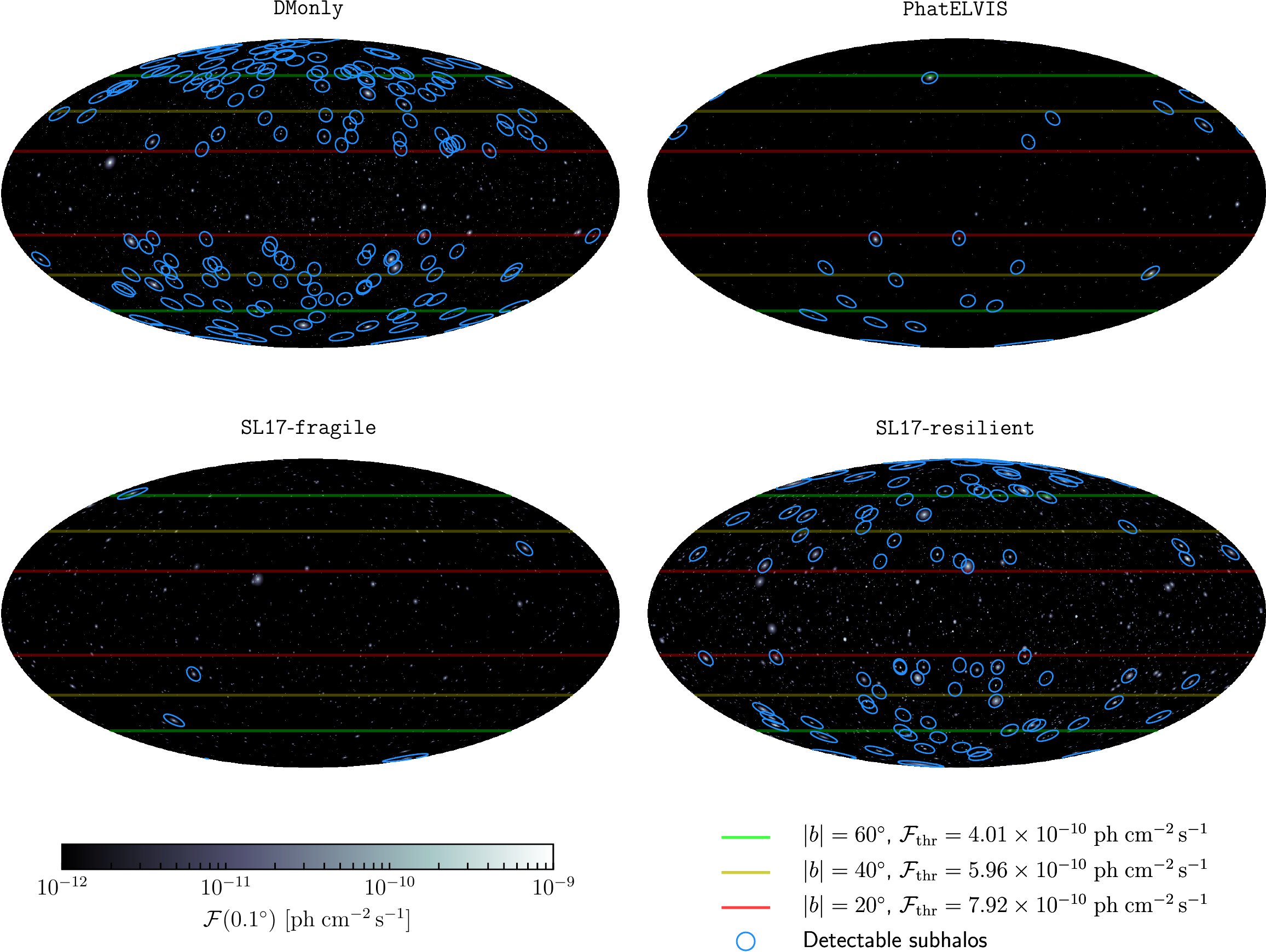}
\caption{For the same realisations as in Fig.~\ref{fig:JvsM_detected}, we display the corresponding all-sky gamma-ray maps of the selected 	halos. Fluxes are computed assuming a DM mass of 100 GeV and an annihilation cross section into $b\bar{b}$ of $5 \times 10^{-24} \, \rm cm^3/s$, for the 3FGL catalogue set-up. We overlay the LAT sensitivity threshold curves at fixed latitude values. The blue circles indicate the subhaloes that are above threshold, and that would therefore be detectable in the 3FGL catalogue.
	Top left: \texttt{DMonly}, top right: \texttt{PhatELVIS}, bottom left: \texttt{SL17-fragile},  bottom right: \texttt{SL17-resilient}. The orange circles indicating the detectable subhaloes have a diameter of $7^\circ$.
	}
\label{fig:skymaps}
\end{figure}

In Fig.~\ref{fig:skymaps}, we display the all-sky gamma-ray maps of selected haloes corresponding to the realisations shown in Fig.~\ref{fig:JvsM}. Fluxes are computed again assuming a DM mass of 100 GeV and an annihilation cross section into $b\bar{b}$ of $5 \times 10^{-24} \, \rm cm^3/s$, for the 3FGL catalogue set-up. Subhaloes whose flux exceeds the sensitivity threshold are highlighted by light blue circles on the skymaps.
Besides the latitude cut $|b|>20^{\circ}$, we can see that bright clumps at high latitude remain undetectable because of the latitude (and DM mass dependence) of the LAT detection threshold. 
We also note that subhaloes can have a very small angular extension on the sky and still be detectable, as can be seen in particular on the \texttt{SL17-resilient} skymap (bottom right). This is due to the cuspy density profile of DM haloes: Even if the structure is stripped off its outer layers by tidal effects, the \Jf~ is only mildly affected and can remain quite high.

\medskip

Dedicated searches for DM subhalo candidates among \Fermi-LAT unassociated sources have been performed in the past through spectral and spatial analyses, often based on machine learning classification algorithms, see the latest analysis in~\cite{Coronado-Blazquez:2019puc}.
The most recent analysis found 16 (4, 24) DM subhalo candidates in the 3FGL (2FHL, 3FHL) catalogue~\cite{Coronado-Blazquez:2019puc}. 
The flux sensitivity threshold inferred from Fig.~9 of~\cite{Coronado-Blazquez:2019puc} are quite similar to the corresponding sensitivity curves of the 2FHL -- so we will provide predictions for the 2FHL in the present work.
Also, the limits from the 3FGL and 2FHL are completely complementary and the strongest over the full DM mass range considered in~\cite{Coronado-Blazquez:2019puc}.

Knowing the number of DM subhalo candidates in \Fermi-LAT catalogues, it is possible to infer an upper limit on the DM annihilation cross section: For a given DM mass, this would be the value of $\sv$ giving a number of detectable subhaloes equal to the number of DM subhalo candidates, $N_{\rm c}$.
The strongest bounds on DM would of course correspond to the case in which $N_{\rm c}=0$. On the other hand, the most conservative limits come from the case where $N_{\rm c}$ is equal to the number of unassociated sources --which is anyhow unrealistic since most likely the largest fraction of these is indeed made up by standard astrophysical objects.

\begin{figure}[!h]
	\centering
	\includegraphics[width=0.49\columnwidth]{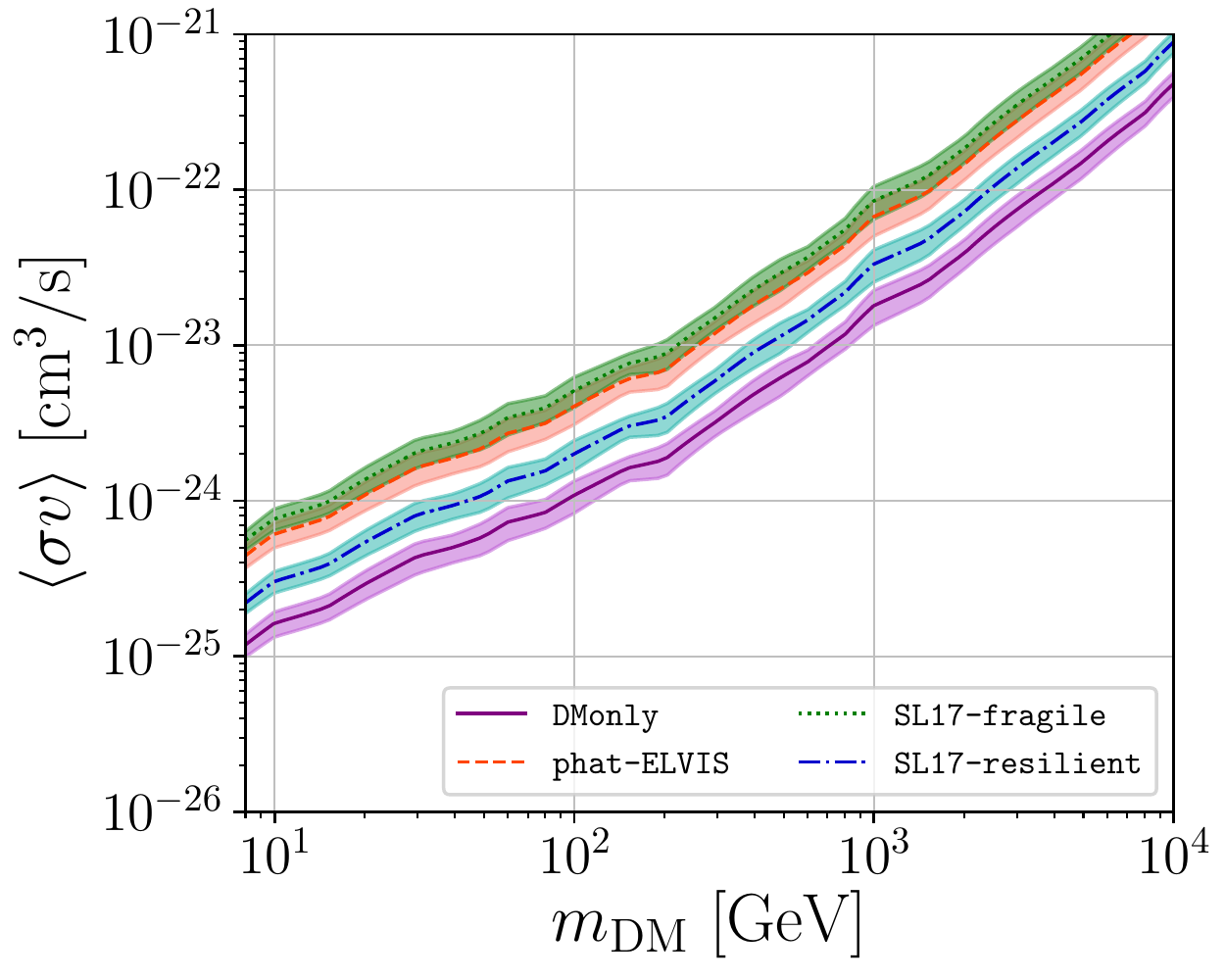}
	\includegraphics[width=0.49\columnwidth]{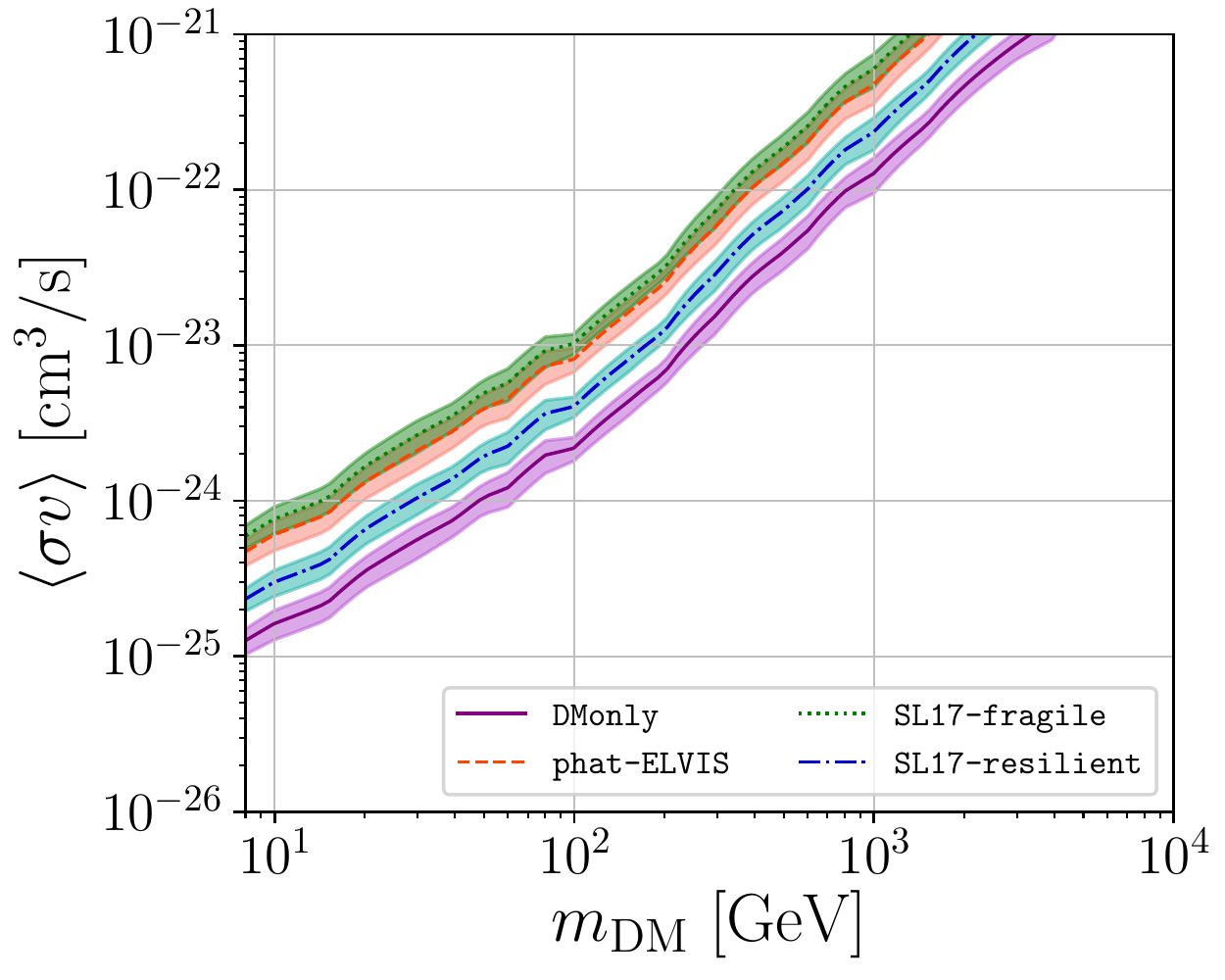}
	\includegraphics[width=0.49\columnwidth]{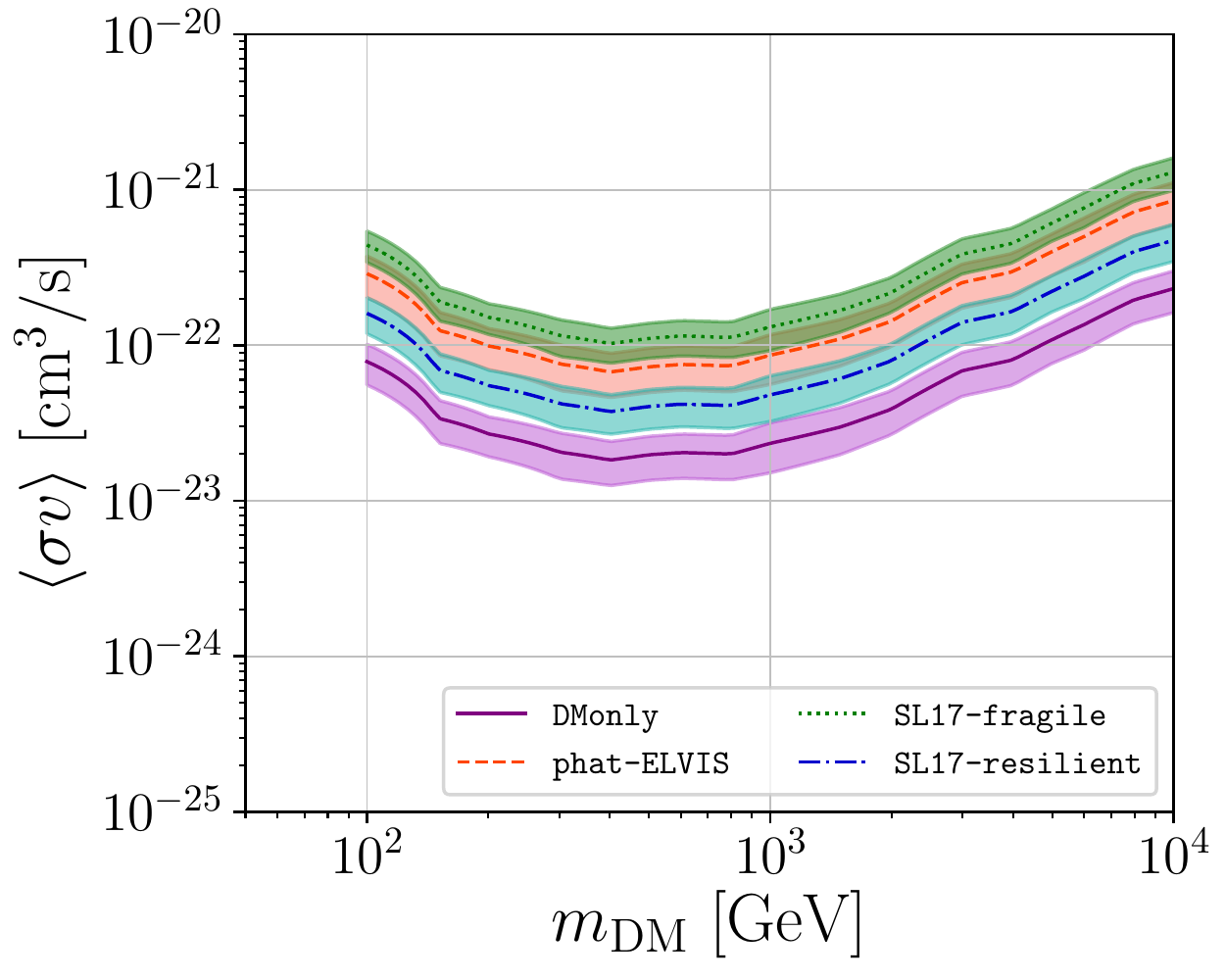}
	\includegraphics[width=0.49\columnwidth]{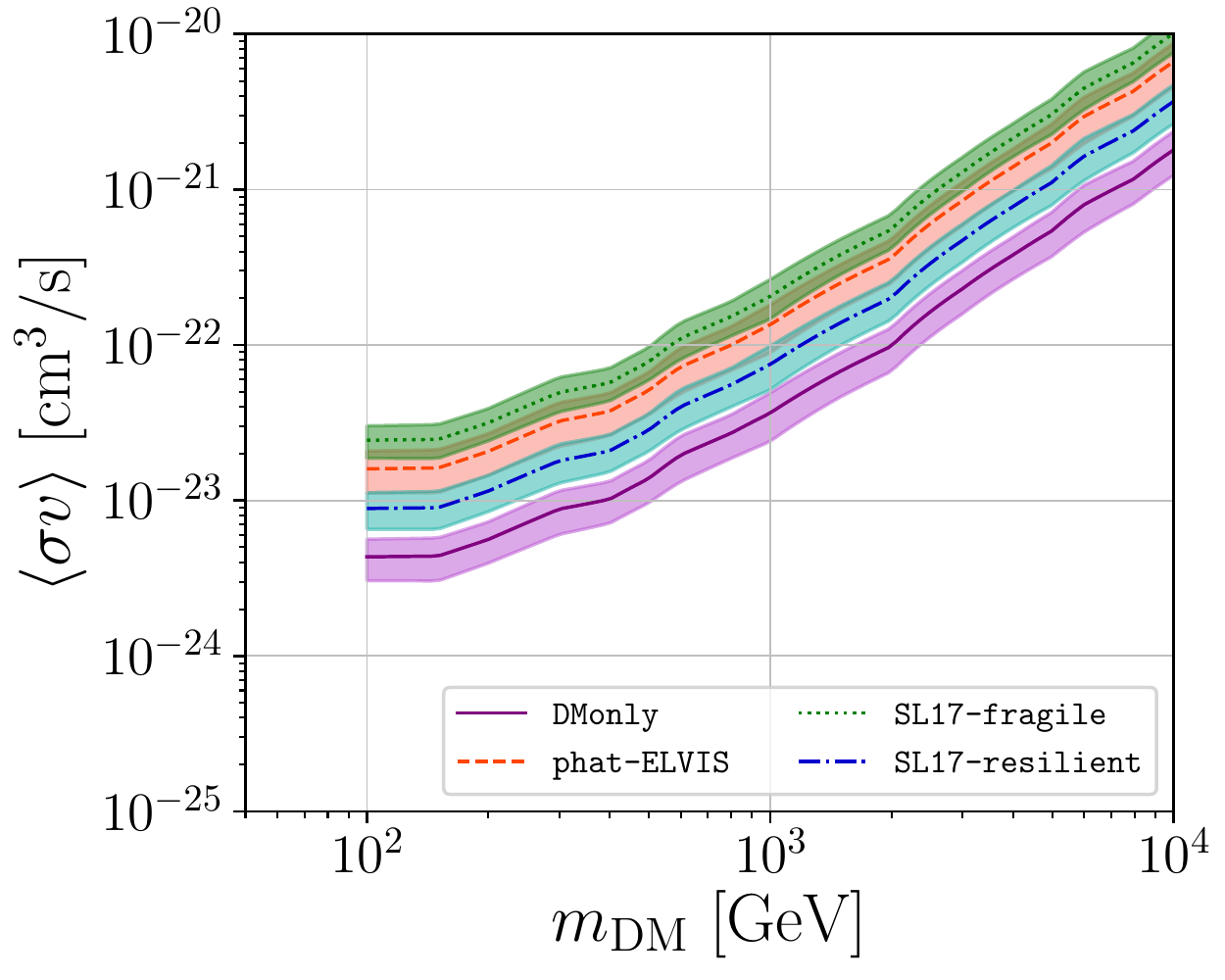}
\caption{
Upper limits on the DM annihilation cross section, $\sv$, from the observation of 16 (4) DM subhalo candidates, $N_{\rm cand}$, in the 3FGL (2FHL) catalogue (the number of candidates is taken from~\cite{Coronado-Blazquez:2019puc}). 
We show the limit for \texttt{DMonly} (purple curve), \texttt{PhatELVIS} (red dashed curve), \texttt{SL17-resilient} (blue dotted-dashed curve) and \texttt{SL17-fragile} (green dotted curve).
The same colour-code applies to uncertainty bands which represent the spread due to the 1000 Monte Carlo realisations for each subhalo model.
	\emph{Top left (right) panel}: Annihilation into $b\bar{b}$ ($\tau^+ \tau^-$) for the 3FGL catalogue.
	\emph{Bottom left (right) panel}: Annihilation into $b\bar{b}$ ($\tau^+ \tau^-$) for the 2FHL catalogue.} 
\label{fig:sv_mass} 
\end{figure}

In Fig.~\ref{fig:sv_mass}, we present upper limits on the DM annihilation cross section as a function of the particle mass that comes from comparing the number of detectable subhaloes in the four models under consideration with the number of DM subhalo candidates from~\cite{Coronado-Blazquez:2019puc}. 
The upper limit on the cross section is defined as the maximum value of $\sv$ for which the predicted subhalo gamma-ray fluxes are equal to the catalogue sensitivity flux threshold. 
The uncertainty bands correspond to the uncertainty in the subhalo modelling, propagating the spread in the 1000 Monte Carlo realisations of the subhalo models (namely, the ``Galactic subhaloes variance).
We find that the \texttt{DMonly} configuration leads to the strongest bounds on the annihilation cross section. The bound from \texttt{SL17-resilient} is factor of $\sim$2 weaker, while the bounds from \texttt{PhatELVIS} and \texttt{SL17-fragile} are similar and are $\sim 5-6$ times weaker. Unsurprisingly, configurations where tidal disruption is not very efficient lead to the strongest bounds. We can compare our bounds from the \texttt{DMonly} model with the limits obtained by \cite{Coronado-Blazquez:2019puc} for the 3FGL and 2FHL catalogues (the authors also computed a limit for the updated 3FHL catalogue). Their limits are a factor of $\sim 3$ stronger for both catalogues.  
At the origin of this difference there can be various reasons: for example, we recall that our \texttt{DMonly} model is based on the \textit{Aquarius} cosmological simulation while the subhalo model used in \cite{Coronado-Blazquez:2019puc} is based on \textit{Via Lactea II} \cite{Diemand:2008in}. subhaloes in these simulations have a different spatial distribution and the total number of resolved objects within the virial radius of the galactic halo also differs, hence there is no reason to expect the exact same gamma-ray prediction from both models. Also,~\cite{Coronado-Blazquez:2019puc} consider \Jf~integration angles equal to $r_s$, while we integrate only up to $0.1^\circ$. 
We note that in the case of the 2FHL our limits are cut at 100 GeV masses; below this mass the limits steeply increase because of a loss of sensitivity of the 2FHL catalogue.

\begin{figure}[!h]
	\centering
	\includegraphics[width=0.49\columnwidth]{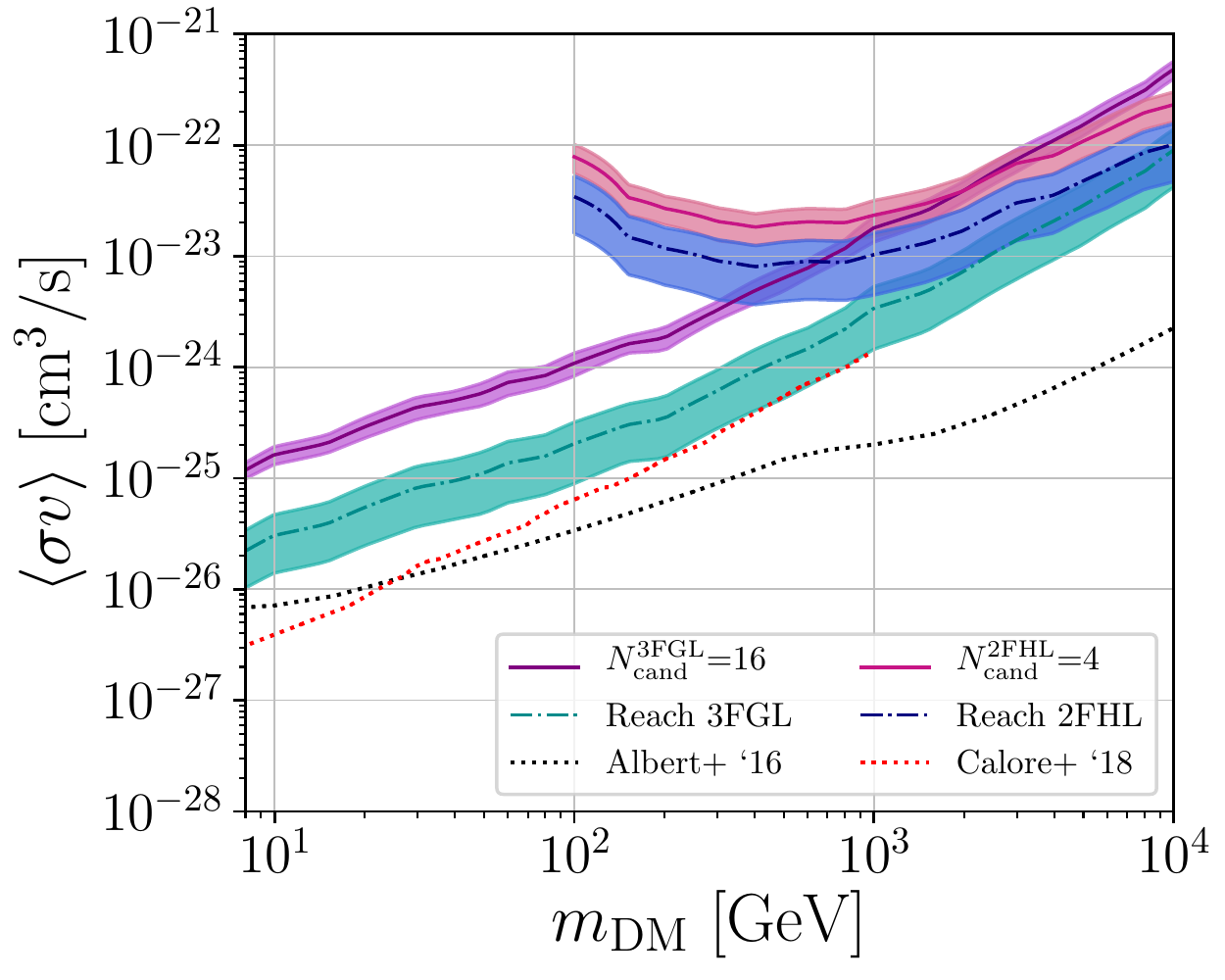}
	\includegraphics[width=0.49\columnwidth]{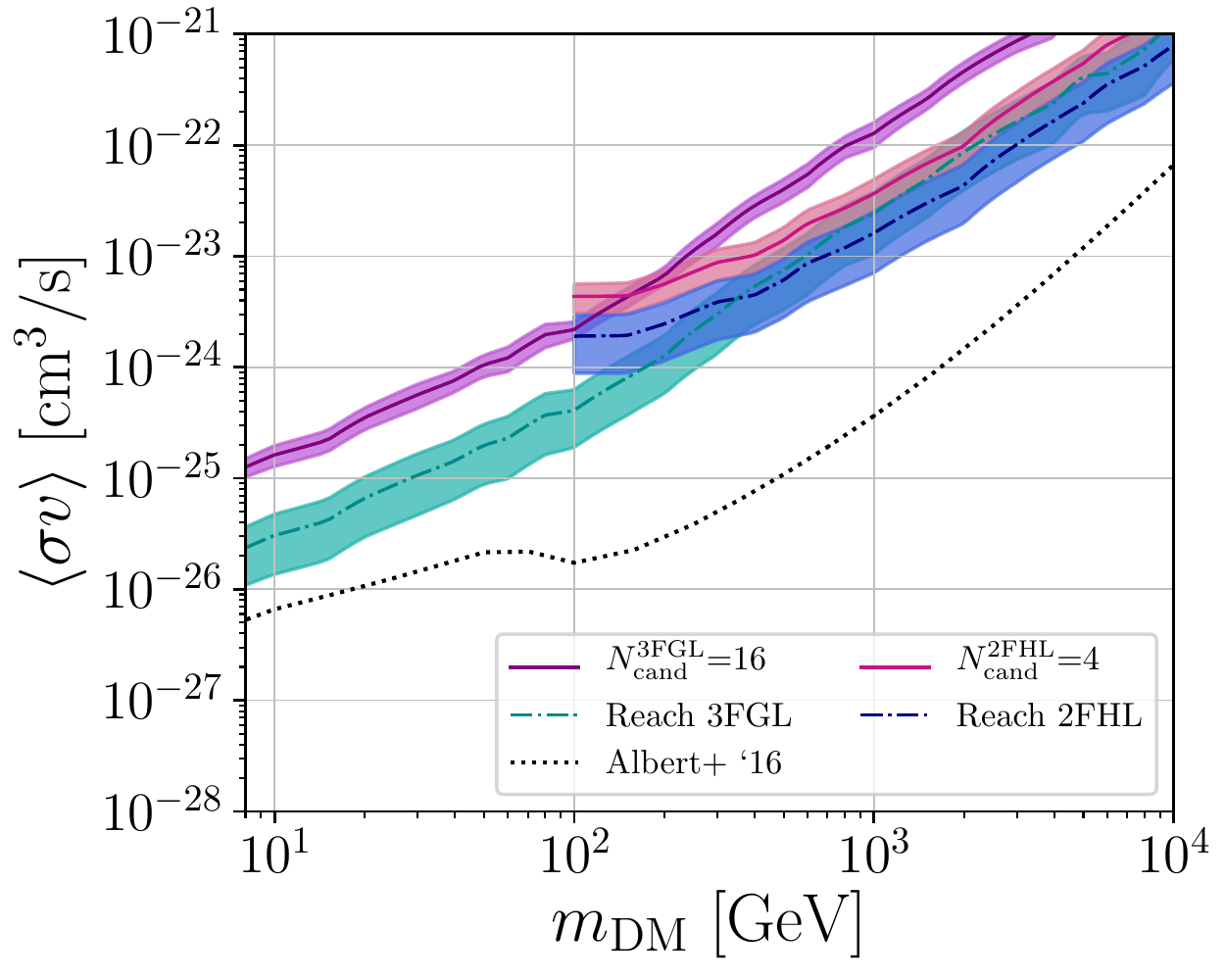}
\caption{Upper limits on the DM annihilation cross section, $\sv$, from the observation of 16 (4) DM subhalo candidates, $N_{\rm cand}$, in the 3FGL --purple solid-- (2FHL --red solid) catalogue, for the \texttt{DMonly}. 
The sensitivity reach ($N_{\rm cand} = 0$) of the 3FGL (2FHL) is also shown by the turquoise dashed-dotted line (blue dashed-dotted line). The same colour-code applies to uncertainty bands which represent the spread due to the 1000 Monte Carlo realisations of the subhalo model. \emph{Left (Right) panel}: Annihilation into $b$-quark ($\tau$ lepton) finale states.
Overlaid, the limits from gamma-ray observations towards dwarf spheroidal galaxies from Albert et al.~2016~\cite{Fermi-LAT:2016uux} (black dotted), and Calore et al.~2018~\cite{Calore:2018sdx} (red dotted).
}
\label{fig:sv_mass_combined} 
\end{figure}

In Fig.~\ref{fig:sv_mass_combined}, we put together the limits from the 3FGL and 2FHL catalogues and compare them to existing limits from gamma-ray observations of dwarf spheroidal galaxies \cite{Fermi-LAT:2016uux,Calore:2018sdx}.
We also display the ``sensitivity reach" of DM searches towards unassociated gamma-ray sources, namely the limit on the annihilation cross section one gets imposing that no DM subhalo candidate remains among unassociated gamma-ray sources in the 3FGL and 2FHL catalogues.
We stress that cutting the integration radius up to 0.1$^\circ$ leads to less strong bounds on the annihilation cross section (about a factor of 2 at all masses). Again, we believe our choice to be truly conservative, against what was done in the past. 
We can therefore see that the limits on the DM parameter space from the dark subhalo search is not as competitive as the search towards dwarf spheroidal galaxies -- at least with present catalogues (and current sensitivity threshold).
Indeed, the sensitivity reach for the 3FGL and 2FHL  catalogues is always above the current limits from dwarf galaxies.

\section{Discussion and conclusions}
\label{sec:conclusions}

In this work we have assessed the detectability of DM subhaloes in \Fermi-LAT catalogues taking into account the uncertainties associated to the modelling of the Galactic subhalo population. We have investigated four different subhalo models: one based on the \textit{Aquarius} DM-only simulation \cite{Springel:2008b}, one on the \textit{Phat-ELVIS} DM simulation which incorporates a disc potential \cite{Kelley:2018pdy}, and two configurations based on an analytical model \cite{Stref:2016uzb}. The incorporation of these models in CLUMPY \cite{2012CoPhC.183..656C,Bonnivard:2015pia,Hutten:2018aix} allowed us to perform 1000 Monte Carlo realisations for each configuration. We then identified among each realisations the detectable subhaloes according to the criterion derived by \citet{Calore:2016ogv} for the 3FGL and 2FHL \textit{Fermi} point-source catalogues. We obtained the DM annihilation cross section required to detect at least one subhalo, see Tab.~\ref{tab:Ndet_sv}, to be a few $\times 10^{-25}$ ($\times 10^{-23}$) for the 3FGL (2FHL) catalogue set-up. We find that, irrespective of the subhalo model, the minimal cross section is already ruled out by gamma-ray observation of dwarf galaxies. Using the unassociated point-sources in the \Fermi-LAT catalogues, we can derive upper limits on the annihilation cross section as a function of the DM mass. We have done so using the number of subhalo candidates found by \citet{Coronado-Blazquez:2019puc} to get a conservative limit and we have shown the result in Fig.~\ref{fig:sv_mass}. A more stringent bound is obtained if we assume all the unassociated sources are in fact explained by conventional astrophysical objects. We show the corresponding bound along with existing limits from dwarf galaxies in Fig.~\ref{fig:sv_mass_combined}. We find that even for the \texttt{DMonly} configuration, which does not include tidal disruption from baryons, the subhalo bound is less stringent than the dwarf galaxy limit.

Comparing the results obtained with the different subhalo configurations, we find that baryonic effects on the subhalo population are significant and lead to DM constraints that are less stringent by a factor of $\sim$2 to $\sim$5. This uncertainty comes from the unknown resilience of DM subhaloes to tidal disruption. 

We note that, compared to previous works, we conservatively adopt a radius of 0.1${^\circ}$ for the \Jf~integration.
This choice is fully consistent with the computation of the \Fermi-LAT threshold to subhalo signals as point-like sourcessubhaloes. 
Unavoidably, this leads to limits on the annihilation cross section which are a factor of a few less stringent than what found in the literature towards dark subhaloes. 
Nevertheless, we mention that stronger constraints can be set by looking at extended \Fermi-LAT unassociated sources. Spatial extension of a gamma-ray unassociated source at high-latitude is generally considered a very promising hint for the DM nature of that emission. So far, however, no source has been flagged as extended~\cite{Coronado-Blazquez:2019puc}, and, in general, only a few subhaloes are found to be extended in galaxy formation simulations~\cite{Calore:2016ogv,Schoonenberg:2016aml}.
The DM subhalo models studied in the present work, instead, predict many subhaloes to have significant angular extension. This means that, in order to properly assess their detectability, the sensitivity to the LAT to such a type of extended signals needs to be computed. We leave this work for a future publication, where we will also derive corresponding constraints on the DM parameter space.

\medskip

In the future, CTA~\cite{Acharya:2013sxa} is expected to boost the search for DM particles with high-energy gamma rays. Sensitivity studies show that we expect at least factor of 10 improvements in the limits from the Galactic center analysis~\cite{Acharya:2017ttl,Silverwood:2014yza}. Also, searches towards dark subhaloes can be competitive, for example, by exploiting the data from a foresseen large-sky survey ~\cite{Acharya:2017ttl,Hutten:2016jko}. In particular, CTA deep follow-up  observations of subhalo candidates or of hints of weak signals in gamma-ray surveys will provide an unprecedented discovery potential for indirect DM signals.
Limits from known dwarf satellites with future telescopes will be very promising also because of the  revolutionary results promised by the Large Synoptic Survey Telescope (LSST)~\cite{2009arXiv0912.0201L}: tens to hundreds new faint satellites of the Milky Way are expected to be discovered and their stellar kinematics to be measured with high accuracy, characterising their DM content. This will further accelerate the DM constraining power of already existing data, such as the ones collected by \Fermi-LAT and future CTA observations~\cite{Ando:2019rvr}.
Finally, lower gamma-ray energies (i.e. <100 MeV) represent an almost unexplored territory. Advanced proposals for MeV telescopes exist~\cite{Moiseev:2015lva,DeAngelis:2016slk}, and future prospects look very promising, offering new opportunities to discover and/or constrain DM particle models~\cite{DeAngelis:2017gra}. 
\vspace{6pt} 

\authorcontributions{Conceptualization, F.C.; Formal analysis, F.C.; Writing--original draft preparation, F.C.; writing--review and editing, all authors.}

\funding{The work of M.H. is supported by the Max Planck society (MPG). M.S. acknowledges support from the ANR project GaDaMa (ANR-18-CE31-0006), the~CNRS IN2P3-Theory/INSU-PNHE-PNCG project ``Galactic Dark Matter'', and~European Union's Horizon 2020 research and innovation program under the Marie Sk\l{}odowska-Curie grant agreements N$^\circ$ 690575 and N$^\circ$ 674896.}

\acknowledgments{We thank D.~Maurin for the reading of the manuscript and constructive feedback, and M.~Di Mauro for useful discussions. We also thank M.~Doro and M.~A.~S{\'a}nchez-Conde for inviting us to contribute to the Special Issue ``The Role of Halo Substructure in Gamma-Ray Dark Matter Searches".}

\conflictsofinterest{The authors declare no conflict of interest. The funders had no role in the design of the study; in the collection, analyses, or interpretation of data; in the writing of the manuscript, or in the decision to publish the results.}

\reftitle{References}


\externalbibliography{yes}
\bibliography{galaxies_subhalos}

\begin{thebibliography}{-------}
\providecommand{\natexlab}[1]{#1}

\bibitem[Aghanim \em{et~al.}(2018)Aghanim et~al.]{Aghanim:2018eyx}
Aghanim, N.; others.
\newblock {Planck 2018 results. VI. Cosmological parameters} {\bf 2018}.
\newblock  \href{http://xxx.lanl.gov/abs/1807.06209}{{\normalfont
  [arXiv:astro-ph.CO/1807.06209]}}.

\bibitem[Bertone \em{et~al.}(2005)Bertone, Hooper, and Silk]{Bertone:2004pz}
Bertone, G.; Hooper, D.; Silk, J.
\newblock {Particle dark matter: Evidence, candidates and constraints}.
\newblock {\em Phys. Rept.} {\bf 2005}, {\em 405},~279--390,
  \href{http://xxx.lanl.gov/abs/hep-ph/0404175}{{\normalfont
  [arXiv:hep-ph/hep-ph/0404175]}}.
\newblock
  doi:{\changeurlcolor{black}\href{https://doi.org/10.1016/j.physrep.2004.08.031}{\detokenize{10.1016/j.physrep.2004.08.031}}}.

\bibitem[Bringmann and Weniger(2012)]{Bringmann:2012ez}
Bringmann, T.; Weniger, C.
\newblock {Gamma Ray Signals from Dark Matter: Concepts, Status and Prospects}.
\newblock {\em Phys. Dark Univ.} {\bf 2012}, {\em 1},~194--217,
  \href{http://xxx.lanl.gov/abs/1208.5481}{{\normalfont
  [arXiv:hep-ph/1208.5481]}}.
\newblock
  doi:{\changeurlcolor{black}\href{https://doi.org/10.1016/j.dark.2012.10.005}{\detokenize{10.1016/j.dark.2012.10.005}}}.

\bibitem[{Gaskins}(2016)]{2016ConPh..57..496G}
{Gaskins}, J.M.
\newblock {A review of indirect searches for particle dark matter}.
\newblock {\em Contemporary Physics} {\bf 2016}, {\em 57},~496--525,
  \href{http://xxx.lanl.gov/abs/1604.00014}{{\normalfont
  [arXiv:astro-ph.HE/1604.00014]}}.
\newblock
  doi:{\changeurlcolor{black}\href{https://doi.org/10.1080/00107514.2016.1175160}{\detokenize{10.1080/00107514.2016.1175160}}}.

\bibitem[Acharya \em{et~al.}(2013)Acharya et~al.]{Acharya:2013sxa}
Acharya, B.S.; others.
\newblock {Introducing the CTA concept}.
\newblock {\em Astropart. Phys.} {\bf 2013}, {\em 43},~3--18.
\newblock
  doi:{\changeurlcolor{black}\href{https://doi.org/10.1016/j.astropartphys.2013.01.007}{\detokenize{10.1016/j.astropartphys.2013.01.007}}}.

\bibitem[Strigari(2018)]{Strigari:2018utn}
Strigari, L.E.
\newblock {Dark matter in dwarf spheroidal galaxies and indirect detection: a
  review}.
\newblock {\em Rept. Prog. Phys.} {\bf 2018}, {\em 81},~056901,
  \href{http://xxx.lanl.gov/abs/1805.05883}{{\normalfont
  [arXiv:astro-ph.CO/1805.05883]}}.
\newblock
  doi:{\changeurlcolor{black}\href{https://doi.org/10.1088/1361-6633/aaae16}{\detokenize{10.1088/1361-6633/aaae16}}}.

\bibitem[Ackermann \em{et~al.}(2015)Ackermann et~al.]{Ackermann:2015zua}
Ackermann, M.; others.
\newblock {Searching for Dark Matter Annihilation from Milky Way Dwarf
  Spheroidal Galaxies with Six Years of Fermi Large Area Telescope Data}.
\newblock {\em Phys. Rev. Lett.} {\bf 2015}, {\em 115},~231301,
  \href{http://xxx.lanl.gov/abs/1503.02641}{{\normalfont
  [arXiv:astro-ph.HE/1503.02641]}}.
\newblock
  doi:{\changeurlcolor{black}\href{https://doi.org/10.1103/PhysRevLett.115.231301}{\detokenize{10.1103/PhysRevLett.115.231301}}}.

\bibitem[Calore \em{et~al.}(2018)Calore, Serpico, and Zaldivar]{Calore:2018sdx}
Calore, F.; Serpico, P.D.; Zaldivar, B.
\newblock {Dark matter constraints from dwarf galaxies: a data-driven
  analysis}.
\newblock {\em JCAP} {\bf 2018}, {\em 1810},~029,
  \href{http://xxx.lanl.gov/abs/1803.05508}{{\normalfont
  [arXiv:astro-ph.HE/1803.05508]}}.
\newblock
  doi:{\changeurlcolor{black}\href{https://doi.org/10.1088/1475-7516/2018/10/029}{\detokenize{10.1088/1475-7516/2018/10/029}}}.

\bibitem[{Springel} \em{et~al.}(2008){Springel}, {Wang}, {Vogelsberger},
  {Ludlow}, {Jenkins}, {Helmi}, {Navarro}, {Frenk}, and
  {White}]{2008MNRAS.391.1685S}
{Springel}, V.; {Wang}, J.; {Vogelsberger}, M.; {Ludlow}, A.; {Jenkins}, A.;
  {Helmi}, A.; {Navarro}, J.F.; {Frenk}, C.S.; {White}, S.D.M.
\newblock {The Aquarius Project: the subhaloes of galactic haloes}.
\newblock {\em \mnras} {\bf 2008}, {\em 391},~1685--1711,
  \href{http://xxx.lanl.gov/abs/0809.0898}{{\normalfont
  [arXiv:astro-ph/0809.0898]}}.
\newblock
  doi:{\changeurlcolor{black}\href{https://doi.org/10.1111/j.1365-2966.2008.14066.x}{\detokenize{10.1111/j.1365-2966.2008.14066.x}}}.

\bibitem[{Diemand} \em{et~al.}(2008){Diemand}, {Kuhlen}, {Madau}, {Zemp},
  {Moore}, {Potter}, and {Stadel}]{2008Natur.454..735D}
{Diemand}, J.; {Kuhlen}, M.; {Madau}, P.; {Zemp}, M.; {Moore}, B.; {Potter},
  D.; {Stadel}, J.
\newblock {Clumps and streams in the local dark matter distribution}.
\newblock {\em \nat} {\bf 2008}, {\em 454},~735--738,
  \href{http://xxx.lanl.gov/abs/0805.1244}{{\normalfont
  [arXiv:astro-ph/0805.1244]}}.
\newblock
  doi:{\changeurlcolor{black}\href{https://doi.org/10.1038/nature07153}{\detokenize{10.1038/nature07153}}}.

\bibitem[{Zavala} and {Frenk}(2019)]{2019Galax...7...81Z}
{Zavala}, J.; {Frenk}, C.S.
\newblock {Dark Matter Haloes and Subhaloes}.
\newblock {\em Galaxies} {\bf 2019}, {\em 7},~81,
  \href{http://xxx.lanl.gov/abs/1907.11775}{{\normalfont
  [arXiv:astro-ph.CO/1907.11775]}}.
\newblock
  doi:{\changeurlcolor{black}\href{https://doi.org/10.3390/galaxies7040081}{\detokenize{10.3390/galaxies7040081}}}.

\bibitem[Acero \em{et~al.}(2015)Acero et~al.]{Acero:2015gva}
Acero, F.; others.
\newblock {Fermi Large Area Telescope Third Source Catalog}.
\newblock {\em Astrophys. J. Suppl.} {\bf 2015}, {\em 218},~23.
\newblock
  doi:{\changeurlcolor{black}\href{https://doi.org/10.1088/0067-0049/218/2/23}{\detokenize{10.1088/0067-0049/218/2/23}}}.

\bibitem[Ackermann \em{et~al.}(2016)Ackermann et~al.]{Ackermann:2015uya}
Ackermann, M.; others.
\newblock {2FHL: The Second Catalog of Hard Fermi-LAT Sources}.
\newblock {\em Astrophys. J. Suppl.} {\bf 2016}, {\em 222},~5,
  \href{http://xxx.lanl.gov/abs/1508.04449}{{\normalfont
  [arXiv:astro-ph.HE/1508.04449]}}.
\newblock
  doi:{\changeurlcolor{black}\href{https://doi.org/10.3847/0067-0049/222/1/5}{\detokenize{10.3847/0067-0049/222/1/5}}}.

\bibitem[Mirabal \em{et~al.}(2016)Mirabal, Charles, Ferrara, Gonthier, Harding,
  Sánchez-Conde, and Thompson]{Mirabal:2016huj}
Mirabal, N.; Charles, E.; Ferrara, E.C.; Gonthier, P.L.; Harding, A.K.;
  Sánchez-Conde, M.A.; Thompson, D.J.
\newblock {3FGL Demographics Outside the Galactic Plane using Supervised
  Machine Learning: Pulsar and Dark Matter Subhalo Interpretations}.
\newblock {\em Astrophys. J.} {\bf 2016}, {\em 825},~69,
  \href{http://xxx.lanl.gov/abs/1605.00711}{{\normalfont
  [arXiv:astro-ph.HE/1605.00711]}}.
\newblock
  doi:{\changeurlcolor{black}\href{https://doi.org/10.3847/0004-637X/825/1/69}{\detokenize{10.3847/0004-637X/825/1/69}}}.

\bibitem[Salvetti \em{et~al.}(2017)Salvetti, Chiaro, La~Mura, and
  Thompson]{Salvetti:2017nkp}
Salvetti, D.; Chiaro, G.; La~Mura, G.; Thompson, D.J.
\newblock {3FGLzoo: classifying 3FGL unassociated Fermi-LAT gamma-ray sources
  by artificial neural networks}.
\newblock {\em Mon. Not. Roy. Astron. Soc.} {\bf 2017}, {\em 470},~1291--1297,
  \href{http://xxx.lanl.gov/abs/1705.09832}{{\normalfont
  [arXiv:astro-ph.HE/1705.09832]}}.
\newblock
  doi:{\changeurlcolor{black}\href{https://doi.org/10.1093/mnras/stx1328}{\detokenize{10.1093/mnras/stx1328}}}.

\bibitem[Coronado-Blazquez \em{et~al.}(2019)Coronado-Blazquez, Sanchez-Conde,
  Dominguez, Aguirre-Santaella, Di~Mauro, Mirabal, Nieto, and
  Charles]{Coronado-Blazquez:2019puc}
Coronado-Blazquez, J.; Sanchez-Conde, M.A.; Dominguez, A.; Aguirre-Santaella,
  A.; Di~Mauro, M.; Mirabal, N.; Nieto, D.; Charles, E.
\newblock {Unidentified Gamma-ray Sources as Targets for Indirect Dark Matter
  Detection with the Fermi-Large Area Telescope}.
\newblock {\em JCAP} {\bf 2019}, {\em 1907},~020,
  \href{http://xxx.lanl.gov/abs/1906.11896}{{\normalfont
  [arXiv:astro-ph.HE/1906.11896]}}.
\newblock
  doi:{\changeurlcolor{black}\href{https://doi.org/10.1088/1475-7516/2019/07/020}{\detokenize{10.1088/1475-7516/2019/07/020}}}.

\bibitem[Belikov \em{et~al.}(2012)Belikov, Hooper, and Buckley]{Belikov:2011pu}
Belikov, A.V.; Hooper, D.; Buckley, M.R.
\newblock {Searching For Dark Matter Subhalos In the Fermi-LAT Second Source
  Catalog}.
\newblock {\em Phys. Rev.} {\bf 2012}, {\em D86},~043504,
  \href{http://xxx.lanl.gov/abs/1111.2613}{{\normalfont
  [arXiv:hep-ph/1111.2613]}}.
\newblock
  doi:{\changeurlcolor{black}\href{https://doi.org/10.1103/PhysRevD.86.043504}{\detokenize{10.1103/PhysRevD.86.043504}}}.

\bibitem[Berlin and Hooper(2014)]{Berlin:2013dva}
Berlin, A.; Hooper, D.
\newblock {Stringent Constraints on the Dark Matter Annihilation Cross Section
  From Subhalo Searches with the Fermi Gamma-Ray Space Telescope}.
\newblock {\em Phys. Rev.} {\bf 2014}, {\em D89},~016014,
  \href{http://xxx.lanl.gov/abs/1309.0525}{{\normalfont
  [arXiv:hep-ph/1309.0525]}}.
\newblock
  doi:{\changeurlcolor{black}\href{https://doi.org/10.1103/PhysRevD.89.016014}{\detokenize{10.1103/PhysRevD.89.016014}}}.

\bibitem[Bertoni \em{et~al.}(2015)Bertoni, Hooper, and Linden]{Bertoni:2015mla}
Bertoni, B.; Hooper, D.; Linden, T.
\newblock {Examining The Fermi-LAT Third Source Catalog In Search Of Dark
  Matter Subhalos}.
\newblock {\em JCAP} {\bf 2015}, {\em 1512},~035,
  \href{http://xxx.lanl.gov/abs/1504.02087}{{\normalfont
  [arXiv:astro-ph.HE/1504.02087]}}.
\newblock
  doi:{\changeurlcolor{black}\href{https://doi.org/10.1088/1475-7516/2015/12/035}{\detokenize{10.1088/1475-7516/2015/12/035}}}.

\bibitem[Schoonenberg \em{et~al.}(2016)Schoonenberg, Gaskins, Bertone, and
  Diemand]{Schoonenberg:2016aml}
Schoonenberg, D.; Gaskins, J.; Bertone, G.; Diemand, J.
\newblock {Dark matter subhalos and unidentified sources in the Fermi 3FGL
  source catalog}.
\newblock {\em JCAP} {\bf 2016}, {\em 1605},~028,
  \href{http://xxx.lanl.gov/abs/1601.06781}{{\normalfont
  [arXiv:astro-ph.HE/1601.06781]}}.
\newblock
  doi:{\changeurlcolor{black}\href{https://doi.org/10.1088/1475-7516/2016/05/028}{\detokenize{10.1088/1475-7516/2016/05/028}}}.

\bibitem[Hooper and Witte(2017)]{Hooper:2016cld}
Hooper, D.; Witte, S.J.
\newblock {Gamma Rays From Dark Matter Subhalos Revisited: Refining the
  Predictions and Constraints}.
\newblock {\em JCAP} {\bf 2017}, {\em 1704},~018,
  \href{http://xxx.lanl.gov/abs/1610.07587}{{\normalfont
  [arXiv:astro-ph.HE/1610.07587]}}.
\newblock
  doi:{\changeurlcolor{black}\href{https://doi.org/10.1088/1475-7516/2017/04/018}{\detokenize{10.1088/1475-7516/2017/04/018}}}.

\bibitem[Calore \em{et~al.}(2017)Calore, De~Romeri, Di~Mauro, Donato, and
  Marinacci]{Calore:2016ogv}
Calore, F.; De~Romeri, V.; Di~Mauro, M.; Donato, F.; Marinacci, F.
\newblock {Realistic estimation for the detectability of dark matter sub-halos
  with Fermi-LAT}.
\newblock {\em Phys. Rev.} {\bf 2017}, {\em D96},~063009,
  \href{http://xxx.lanl.gov/abs/1611.03503}{{\normalfont
  [arXiv:astro-ph.HE/1611.03503]}}.
\newblock
  doi:{\changeurlcolor{black}\href{https://doi.org/10.1103/PhysRevD.96.063009}{\detokenize{10.1103/PhysRevD.96.063009}}}.

\bibitem[Hütten \em{et~al.}(2019)Hütten, Stref, Combet, Lavalle, and
  Maurin]{Hutten:2019tew}
Hütten, M.; Stref, M.; Combet, C.; Lavalle, J.; Maurin, D.
\newblock {$\gamma$-ray and $\nu$ Searches for Dark-Matter Subhalos in the
  Milky Way with a Baryonic Potential}.
\newblock {\em Galaxies} {\bf 2019}, {\em 7},~60,
  \href{http://xxx.lanl.gov/abs/1904.10935}{{\normalfont
  [arXiv:astro-ph.HE/1904.10935]}}.
\newblock
  doi:{\changeurlcolor{black}\href{https://doi.org/10.3390/galaxies7020060}{\detokenize{10.3390/galaxies7020060}}}.

\bibitem[Green \em{et~al.}(2005)Green, Hofmann, and Schwarz]{Green:2005fa}
Green, A.M.; Hofmann, S.; Schwarz, D.J.
\newblock {The First wimpy halos}.
\newblock {\em JCAP} {\bf 2005}, {\em 0508},~003,
  \href{http://xxx.lanl.gov/abs/astro-ph/0503387}{{\normalfont
  [arXiv:astro-ph/astro-ph/0503387]}}.
\newblock
  doi:{\changeurlcolor{black}\href{https://doi.org/10.1088/1475-7516/2005/08/003}{\detokenize{10.1088/1475-7516/2005/08/003}}}.

\bibitem[Bringmann(2009)]{Bringmann:2009vf}
Bringmann, T.
\newblock {Particle Models and the Small-Scale Structure of Dark Matter}.
\newblock {\em New J. Phys.} {\bf 2009}, {\em 11},~105027,
  \href{http://xxx.lanl.gov/abs/0903.0189}{{\normalfont
  [arXiv:astro-ph.CO/0903.0189]}}.
\newblock
  doi:{\changeurlcolor{black}\href{https://doi.org/10.1088/1367-2630/11/10/105027}{\detokenize{10.1088/1367-2630/11/10/105027}}}.

\bibitem[Diemand \em{et~al.}(2008)Diemand, Kuhlen, Madau, Zemp, Moore, Potter,
  and Stadel]{Diemand:2008in}
Diemand, J.; Kuhlen, M.; Madau, P.; Zemp, M.; Moore, B.; Potter, D.; Stadel, J.
\newblock {Clumps and streams in the local dark matter distribution}.
\newblock {\em Nature} {\bf 2008}, {\em 454},~735--738,
  \href{http://xxx.lanl.gov/abs/0805.1244}{{\normalfont
  [arXiv:astro-ph/0805.1244]}}.
\newblock
  doi:{\changeurlcolor{black}\href{https://doi.org/10.1038/nature07153}{\detokenize{10.1038/nature07153}}}.

\bibitem[{Springel} \em{et~al.}(2008){Springel}, {Wang}, {Vogelsberger},
  {Ludlow}, {Jenkins}, {Helmi}, {Navarro}, {Frenk}, and
  {White}]{Springel:2008b}
{Springel}, V.; {Wang}, J.; {Vogelsberger}, M.; {Ludlow}, A.; {Jenkins}, A.;
  {Helmi}, A.; {Navarro}, J.F.; {Frenk}, C.S.; {White}, S.D.M.
\newblock {The Aquarius Project: the subhaloes of galactic haloes}.
\newblock {\em \mnras} {\bf 2008}, {\em 391},~1685--1711.

\bibitem[Moliné \em{et~al.}(2017)Moliné, Sánchez-Conde, Palomares-Ruiz, and
  Prada]{Moline:2016pbm}
Moliné, A.; Sánchez-Conde, M.A.; Palomares-Ruiz, S.; Prada, F.
\newblock {Characterization of subhalo structural properties and implications
  for dark matter annihilation signals}.
\newblock {\em Mon. Not. Roy. Astron. Soc.} {\bf 2017}, {\em 466},~4974--4990,
  \href{http://xxx.lanl.gov/abs/1603.04057}{{\normalfont
  [arXiv:astro-ph.CO/1603.04057]}}.
\newblock
  doi:{\changeurlcolor{black}\href{https://doi.org/10.1093/mnras/stx026}{\detokenize{10.1093/mnras/stx026}}}.

\bibitem[Kelley \em{et~al.}(2018)Kelley, Bullock, Garrison-Kimmel,
  Boylan-Kolchin, Pawlowski, and Graus]{Kelley:2018pdy}
Kelley, T.; Bullock, J.S.; Garrison-Kimmel, S.; Boylan-Kolchin, M.; Pawlowski,
  M.S.; Graus, A.S.
\newblock {Phat ELVIS: The inevitable effect of the Milky Way's disk on its
  dark matter subhaloes} {\bf 2018}.
\newblock  \href{http://xxx.lanl.gov/abs/1811.12413}{{\normalfont
  [arXiv:astro-ph.GA/1811.12413]}}.

\bibitem[Stref and Lavalle(2017)]{Stref:2016uzb}
Stref, M.; Lavalle, J.
\newblock {Modeling dark matter subhalos in a constrained galaxy: Global mass
  and boosted annihilation profiles}.
\newblock {\em Phys. Rev.} {\bf 2017}, {\em D95},~063003,
  \href{http://xxx.lanl.gov/abs/1610.02233}{{\normalfont
  [arXiv:astro-ph.CO/1610.02233]}}.
\newblock
  doi:{\changeurlcolor{black}\href{https://doi.org/10.1103/PhysRevD.95.063003}{\detokenize{10.1103/PhysRevD.95.063003}}}.

\bibitem[van~den Bosch \em{et~al.}(2018)van~den Bosch, Ogiya, Hahn, and
  Burkert]{vandenBosch:2017ynq}
van~den Bosch, F.C.; Ogiya, G.; Hahn, O.; Burkert, A.
\newblock {Disruption of Dark Matter Substructure: Fact or Fiction?}
\newblock {\em Mon. Not. Roy. Astron. Soc.} {\bf 2018}, {\em 474},~3043--3066,
  \href{http://xxx.lanl.gov/abs/1711.05276}{{\normalfont
  [arXiv:astro-ph.GA/1711.05276]}}.
\newblock
  doi:{\changeurlcolor{black}\href{https://doi.org/10.1093/mnras/stx2956}{\detokenize{10.1093/mnras/stx2956}}}.

\bibitem[van~den Bosch and Ogiya(2018)]{vandenBosch:2018tyt}
van~den Bosch, F.C.; Ogiya, G.
\newblock {Dark Matter Substructure in Numerical Simulations: A Tale of
  Discreteness Noise, Runaway Instabilities, and Artificial Disruption}.
\newblock {\em Mon. Not. Roy. Astron. Soc.} {\bf 2018}, {\em 475},~4066--4087,
  \href{http://xxx.lanl.gov/abs/1801.05427}{{\normalfont
  [arXiv:astro-ph.GA/1801.05427]}}.
\newblock
  doi:{\changeurlcolor{black}\href{https://doi.org/10.1093/mnras/sty084}{\detokenize{10.1093/mnras/sty084}}}.

\bibitem[{Errani} and {Pe{\~n}arrubia}(2019)]{2019arXiv190601642E}
{Errani}, R.; {Pe{\~n}arrubia}, J.
\newblock {Can tides disrupt cold dark matter subhaloes?}
\newblock {\em arXiv e-prints} {\bf 2019}, p. arXiv:1906.01642,
  \href{http://xxx.lanl.gov/abs/1906.01642}{{\normalfont
  [arXiv:astro-ph.GA/1906.01642]}}.

\bibitem[Stref \em{et~al.}(2019)Stref, Lacroix, and Lavalle]{galaxies7020065}
Stref, M.; Lacroix, T.; Lavalle, J.
\newblock Remnants of Galactic Subhalos and Their Impact on Indirect
  Dark-Matter Searches.
\newblock {\em Galaxies} {\bf 2019}, {\em 7}.
\newblock
  doi:{\changeurlcolor{black}\href{https://doi.org/10.3390/galaxies7020065}{\detokenize{10.3390/galaxies7020065}}}.

\bibitem[{Charbonnier} \em{et~al.}(2012){Charbonnier}, {Combet}, and
  {Maurin}]{2012CoPhC.183..656C}
{Charbonnier}, A.; {Combet}, C.; {Maurin}, D.
\newblock {CLUMPY: A code for gamma-ray signals from dark matter structures}.
\newblock {\em Computer Physics Communications} {\bf 2012}, {\em
  183},~656--668,  \href{http://xxx.lanl.gov/abs/1201.4728}{{\normalfont
  [arXiv:astro-ph.HE/1201.4728]}}.
\newblock
  doi:{\changeurlcolor{black}\href{https://doi.org/10.1016/j.cpc.2011.10.017}{\detokenize{10.1016/j.cpc.2011.10.017}}}.

\bibitem[Bonnivard \em{et~al.}(2016)Bonnivard, Hütten, Nezri, Charbonnier,
  Combet, and Maurin]{Bonnivard:2015pia}
Bonnivard, V.; Hütten, M.; Nezri, E.; Charbonnier, A.; Combet, C.; Maurin, D.
\newblock {CLUMPY : Jeans analysis, gamma-ray and neutrino fluxes from dark
  matter (sub-)structures}.
\newblock {\em Comput. Phys. Commun.} {\bf 2016}, {\em 200},~336--349,
  \href{http://xxx.lanl.gov/abs/1506.07628}{{\normalfont
  [arXiv:astro-ph.CO/1506.07628]}}.
\newblock
  doi:{\changeurlcolor{black}\href{https://doi.org/10.1016/j.cpc.2015.11.012}{\detokenize{10.1016/j.cpc.2015.11.012}}}.

\bibitem[Hütten \em{et~al.}(2019)Hütten, Combet, and Maurin]{Hutten:2018aix}
Hütten, M.; Combet, C.; Maurin, D.
\newblock {CLUMPY v3: gamma-ray and neutrino signals from dark matter at all
  scales}.
\newblock {\em Comput. Phys. Commun.} {\bf 2019}, {\em 235},~336--345,
  \href{http://xxx.lanl.gov/abs/1806.08639}{{\normalfont
  [arXiv:astro-ph.CO/1806.08639]}}.
\newblock
  doi:{\changeurlcolor{black}\href{https://doi.org/10.1016/j.cpc.2018.10.001}{\detokenize{10.1016/j.cpc.2018.10.001}}}.

\bibitem[Cirelli \em{et~al.}(2011)Cirelli, Corcella, Hektor, Hutsi, Kadastik,
  Panci, Raidal, Sala, and Strumia]{Cirelli:2010xx}
Cirelli, M.; Corcella, G.; Hektor, A.; Hutsi, G.; Kadastik, M.; Panci, P.;
  Raidal, M.; Sala, F.; Strumia, A.
\newblock {PPPC 4 DM ID: A Poor Particle Physicist Cookbook for Dark Matter
  Indirect Detection}.
\newblock {\em JCAP} {\bf 2011}, {\em 1103},~051,
  \href{http://xxx.lanl.gov/abs/1012.4515}{{\normalfont
  [arXiv:hep-ph/1012.4515]}}.
\newblock [Erratum: JCAP1210,E01(2012)],
  doi:{\changeurlcolor{black}\href{https://doi.org/10.1088/1475-7516/2012/10/E01,
  10.1088/1475-7516/2011/03/051}{\detokenize{10.1088/1475-7516/2012/10/E01,
  10.1088/1475-7516/2011/03/051}}}.

\bibitem[Albert \em{et~al.}(2017)Albert et~al.]{Fermi-LAT:2016uux}
Albert, A.; others.
\newblock {Searching for Dark Matter Annihilation in Recently Discovered Milky
  Way Satellites with Fermi-LAT}.
\newblock {\em Astrophys. J.} {\bf 2017}, {\em 834},~110,
  \href{http://xxx.lanl.gov/abs/1611.03184}{{\normalfont
  [arXiv:astro-ph.HE/1611.03184]}}.
\newblock
  doi:{\changeurlcolor{black}\href{https://doi.org/10.3847/1538-4357/834/2/110}{\detokenize{10.3847/1538-4357/834/2/110}}}.

\bibitem[Bonnivard \em{et~al.}(2015)Bonnivard, Combet, Maurin, and
  Walker]{Bonnivard:2014kza}
Bonnivard, V.; Combet, C.; Maurin, D.; Walker, M.G.
\newblock {Spherical Jeans analysis for dark matter indirect detection in dwarf
  spheroidal galaxies - Impact of physical parameters and triaxiality}.
\newblock {\em Mon. Not. Roy. Astron. Soc.} {\bf 2015}, {\em 446},~3002--3021,
  \href{http://xxx.lanl.gov/abs/1407.7822}{{\normalfont
  [arXiv:astro-ph.HE/1407.7822]}}.
\newblock
  doi:{\changeurlcolor{black}\href{https://doi.org/10.1093/mnras/stu2296}{\detokenize{10.1093/mnras/stu2296}}}.

\bibitem[Klop \em{et~al.}(2017)Klop, Zandanel, Hayashi, and Ando]{Klop:2016lug}
Klop, N.; Zandanel, F.; Hayashi, K.; Ando, S.
\newblock {Impact of axisymmetric mass models for dwarf spheroidal galaxies on
  indirect dark matter searches}.
\newblock {\em Phys. Rev.} {\bf 2017}, {\em D95},~123012,
  \href{http://xxx.lanl.gov/abs/1609.03509}{{\normalfont
  [arXiv:astro-ph.CO/1609.03509]}}.
\newblock
  doi:{\changeurlcolor{black}\href{https://doi.org/10.1103/PhysRevD.95.123012}{\detokenize{10.1103/PhysRevD.95.123012}}}.

\bibitem[Ullio and Valli(2016)]{Ullio:2016kvy}
Ullio, P.; Valli, M.
\newblock {A critical reassessment of particle Dark Matter limits from dwarf
  satellites}.
\newblock {\em JCAP} {\bf 2016}, {\em 1607},~025,
  \href{http://xxx.lanl.gov/abs/1603.07721}{{\normalfont
  [arXiv:astro-ph.GA/1603.07721]}}.
\newblock
  doi:{\changeurlcolor{black}\href{https://doi.org/10.1088/1475-7516/2016/07/025}{\detokenize{10.1088/1475-7516/2016/07/025}}}.

\bibitem[Zhu \em{et~al.}(2016)Zhu, Marinacci, Maji, Li, Springel, and
  Hernquist]{Zhu:2015jwa}
Zhu, Q.; Marinacci, F.; Maji, M.; Li, Y.; Springel, V.; Hernquist, L.
\newblock {Baryonic impact on the dark matter distribution in Milky Way-sized
  galaxies and their satellites}.
\newblock {\em Mon. Not. Roy. Astron. Soc.} {\bf 2016}, {\em 458},~1559--1580,
  \href{http://xxx.lanl.gov/abs/1506.05537}{{\normalfont
  [arXiv:astro-ph.CO/1506.05537]}}.
\newblock
  doi:{\changeurlcolor{black}\href{https://doi.org/10.1093/mnras/stw374}{\detokenize{10.1093/mnras/stw374}}}.

\bibitem[Acharya \em{et~al.}(2018)Acharya et~al.]{Acharya:2017ttl}
Acharya, B.S.; others.
\newblock {\em {Science with the Cherenkov Telescope Array}}; WSP,  2018;
  \href{http://xxx.lanl.gov/abs/1709.07997}{{\normalfont
  [arXiv:astro-ph.IM/1709.07997]}}.
\newblock
  doi:{\changeurlcolor{black}\href{https://doi.org/10.1142/10986}{\detokenize{10.1142/10986}}}.

\bibitem[Silverwood \em{et~al.}(2015)Silverwood, Weniger, Scott, and
  Bertone]{Silverwood:2014yza}
Silverwood, H.; Weniger, C.; Scott, P.; Bertone, G.
\newblock {A realistic assessment of the CTA sensitivity to dark matter
  annihilation}.
\newblock {\em JCAP} {\bf 2015}, {\em 1503},~055,
  \href{http://xxx.lanl.gov/abs/1408.4131}{{\normalfont
  [arXiv:astro-ph.HE/1408.4131]}}.
\newblock
  doi:{\changeurlcolor{black}\href{https://doi.org/10.1088/1475-7516/2015/03/055}{\detokenize{10.1088/1475-7516/2015/03/055}}}.

\bibitem[Hütten \em{et~al.}(2016)Hütten, Combet, Maier, and
  Maurin]{Hutten:2016jko}
Hütten, M.; Combet, C.; Maier, G.; Maurin, D.
\newblock {Dark matter substructure modelling and sensitivity of the Cherenkov
  Telescope Array to Galactic dark halos}.
\newblock {\em JCAP} {\bf 2016}, {\em 1609},~047,
  \href{http://xxx.lanl.gov/abs/1606.04898}{{\normalfont
  [arXiv:astro-ph.HE/1606.04898]}}.
\newblock
  doi:{\changeurlcolor{black}\href{https://doi.org/10.1088/1475-7516/2016/09/047}{\detokenize{10.1088/1475-7516/2016/09/047}}}.

\bibitem[{LSST Science Collaboration} \em{et~al.}(2009){LSST Science
  Collaboration}, {Abell}, {Allison}, {Anderson}, {Andrew}, {Angel}, {Armus},
  {Arnett}, {Asztalos}, {Axelrod}, {Bailey}, {Ballantyne}, {Bankert},
  {Barkhouse}, {Barr}, {Barrientos}, {Barth}, {Bartlett}, {Becker}, {Becla},
  {Beers}, {Bernstein}, {Biswas}, {Blanton}, {Bloom}, {Bochanski}, {Boeshaar},
  {Borne}, {Bradac}, {Brandt}, {Bridge}, {Brown}, {Brunner}, {Bullock},
  {Burgasser}, {Burge}, {Burke}, {Cargile}, {Chand rasekharan}, {Chartas},
  {Chesley}, {Chu}, {Cinabro}, {Claire}, {Claver}, {Clowe}, {Connolly}, {Cook},
  {Cooke}, {Cooray}, {Covey}, {Culliton}, {de Jong}, {de Vries}, {Debattista},
  {Delgado}, {Dell'Antonio}, {Dhital}, {Di Stefano}, {Dickinson}, {Dilday},
  {Djorgovski}, {Dobler}, {Donalek}, {Dubois-Felsmann}, {Durech},
  {Eliasdottir}, {Eracleous}, {Eyer}, {Falco}, {Fan}, {Fassnacht}, {Ferguson},
  {Fernandez}, {Fields}, {Finkbeiner}, {Figueroa}, {Fox}, {Francke}, {Frank},
  {Frieman}, {Fromenteau}, {Furqan}, {Galaz}, {Gal-Yam}, {Garnavich},
  {Gawiser}, {Geary}, {Gee}, {Gibson}, {Gilmore}, {Grace}, {Green}, {Gressler},
  {Grillmair}, {Habib}, {Haggerty}, {Hamuy}, {Harris}, {Hawley}, {Heavens},
  {Hebb}, {Henry}, {Hileman}, {Hilton}, {Hoadley}, {Holberg}, {Holman},
  {Howell}, {Infante}, {Ivezic}, {Jacoby}, {Jain}, {R}, {Jedicke}, {Jee},
  {Garrett Jernigan}, {Jha}, {Johnston}, {Jones}, {Juric}, {Kaasalainen},
  {Styliani}, {Kafka}, {Kahn}, {Kaib}, {Kalirai}, {Kantor}, {Kasliwal},
  {Keeton}, {Kessler}, {Knezevic}, {Kowalski}, {Krabbendam}, {Krughoff},
  {Kulkarni}, {Kuhlman}, {Lacy}, {Lepine}, {Liang}, {Lien}, {Lira}, {Long},
  {Lorenz}, {Lotz}, {Lupton}, {Lutz}, {Macri}, {Mahabal}, {Mandelbaum},
  {Marshall}, {May}, {McGehee}, {Meadows}, {Meert}, {Milani}, {Miller},
  {Miller}, {Mills}, {Minniti}, {Monet}, {Mukadam}, {Nakar}, {Neill}, {Newman},
  {Nikolaev}, {Nordby}, {O'Connor}, {Oguri}, {Oliver}, {Olivier}, {Olsen},
  {Olsen}, {Olszewski}, {Oluseyi}, {Padilla}, {Parker}, {Pepper}, {Peterson},
  {Petry}, {Pinto}, {Pizagno}, {Popescu}, {Prsa}, {Radcka}, {Raddick},
  {Rasmussen}, {Rau}, {Rho}, {Rhoads}, {Richards}, {Ridgway}, {Robertson},
  {Roskar}, {Saha}, {Sarajedini}, {Scannapieco}, {Schalk}, {Schindler},
  {Schmidt}, {Schmidt}, {Schneider}, {Schumacher}, {Scranton}, {Sebag},
  {Seppala}, {Shemmer}, {Simon}, {Sivertz}, {Smith}, {Allyn Smith}, {Smith},
  {Spitz}, {Stanford}, {Stassun}, {Strader}, {Strauss}, {Stubbs}, {Sweeney},
  {Szalay}, {Szkody}, {Takada}, {Thorman}, {Trilling}, {Trimble}, {Tyson}, {Van
  Berg}, {Vand en Berk}, {VanderPlas}, {Verde}, {Vrsnak}, {Walkowicz}, {Wand
  elt}, {Wang}, {Wang}, {Warner}, {Wechsler}, {West}, {Wiecha}, {Williams},
  {Willman}, {Wittman}, {Wolff}, {Wood-Vasey}, {Wozniak}, {Young}, {Zentner},
  and {Zhan}]{2009arXiv0912.0201L}
{LSST Science Collaboration}.; {Abell}, P.A.; {Allison}, J.; {Anderson}, S.F.;
  {Andrew}, J.R.; {Angel}, J.R.P.; {Armus}, L.; {Arnett}, D.; {Asztalos}, S.J.;
  {Axelrod}, T.S.; {Bailey}, S.; {Ballantyne}, D.R.; {Bankert}, J.R.;
  {Barkhouse}, W.A.; {Barr}, J.D.; {Barrientos}, L.F.; {Barth}, A.J.;
  {Bartlett}, J.G.; {Becker}, A.C.; {Becla}, J.; {Beers}, T.C.; {Bernstein},
  J.P.; {Biswas}, R.; {Blanton}, M.R.; {Bloom}, J.S.; {Bochanski}, J.J.;
  {Boeshaar}, P.; {Borne}, K.D.; {Bradac}, M.; {Brandt}, W.N.; {Bridge}, C.R.;
  {Brown}, M.E.; {Brunner}, R.J.; {Bullock}, J.S.; {Burgasser}, A.J.; {Burge},
  J.H.; {Burke}, D.L.; {Cargile}, P.A.; {Chand rasekharan}, S.; {Chartas}, G.;
  {Chesley}, S.R.; {Chu}, Y.H.; {Cinabro}, D.; {Claire}, M.W.; {Claver}, C.F.;
  {Clowe}, D.; {Connolly}, A.J.; {Cook}, K.H.; {Cooke}, J.; {Cooray}, A.;
  {Covey}, K.R.; {Culliton}, C.S.; {de Jong}, R.; {de Vries}, W.H.;
  {Debattista}, V.P.; {Delgado}, F.; {Dell'Antonio}, I.P.; {Dhital}, S.; {Di
  Stefano}, R.; {Dickinson}, M.; {Dilday}, B.; {Djorgovski}, S.G.; {Dobler},
  G.; {Donalek}, C.; {Dubois-Felsmann}, G.; {Durech}, J.; {Eliasdottir}, A.;
  {Eracleous}, M.; {Eyer}, L.; {Falco}, E.E.; {Fan}, X.; {Fassnacht}, C.D.;
  {Ferguson}, H.C.; {Fernandez}, Y.R.; {Fields}, B.D.; {Finkbeiner}, D.;
  {Figueroa}, E.E.; {Fox}, D.B.; {Francke}, H.; {Frank}, J.S.; {Frieman}, J.;
  {Fromenteau}, S.; {Furqan}, M.; {Galaz}, G.; {Gal-Yam}, A.; {Garnavich}, P.;
  {Gawiser}, E.; {Geary}, J.; {Gee}, P.; {Gibson}, R.R.; {Gilmore}, K.;
  {Grace}, E.A.; {Green}, R.F.; {Gressler}, W.J.; {Grillmair}, C.J.; {Habib},
  S.; {Haggerty}, J.S.; {Hamuy}, M.; {Harris}, A.W.; {Hawley}, S.L.; {Heavens},
  A.F.; {Hebb}, L.; {Henry}, T.J.; {Hileman}, E.; {Hilton}, E.J.; {Hoadley},
  K.; {Holberg}, J.B.; {Holman}, M.J.; {Howell}, S.B.; {Infante}, L.; {Ivezic},
  Z.; {Jacoby}, S.H.; {Jain}, B.; {R}.; {Jedicke}.; {Jee}, M.J.; {Garrett
  Jernigan}, J.; {Jha}, S.W.; {Johnston}, K.V.; {Jones}, R.L.; {Juric}, M.;
  {Kaasalainen}, M.; {Styliani}.; {Kafka}.; {Kahn}, S.M.; {Kaib}, N.A.;
  {Kalirai}, J.; {Kantor}, J.; {Kasliwal}, M.M.; {Keeton}, C.R.; {Kessler}, R.;
  {Knezevic}, Z.; {Kowalski}, A.; {Krabbendam}, V.L.; {Krughoff}, K.S.;
  {Kulkarni}, S.; {Kuhlman}, S.; {Lacy}, M.; {Lepine}, S.; {Liang}, M.; {Lien},
  A.; {Lira}, P.; {Long}, K.S.; {Lorenz}, S.; {Lotz}, J.M.; {Lupton}, R.H.;
  {Lutz}, J.; {Macri}, L.M.; {Mahabal}, A.A.; {Mandelbaum}, R.; {Marshall}, P.;
  {May}, M.; {McGehee}, P.M.; {Meadows}, B.T.; {Meert}, A.; {Milani}, A.;
  {Miller}, C.J.; {Miller}, M.; {Mills}, D.; {Minniti}, D.; {Monet}, D.;
  {Mukadam}, A.S.; {Nakar}, E.; {Neill}, D.R.; {Newman}, J.A.; {Nikolaev}, S.;
  {Nordby}, M.; {O'Connor}, P.; {Oguri}, M.; {Oliver}, J.; {Olivier}, S.S.;
  {Olsen}, J.K.; {Olsen}, K.; {Olszewski}, E.W.; {Oluseyi}, H.; {Padilla},
  N.D.; {Parker}, A.; {Pepper}, J.; {Peterson}, J.R.; {Petry}, C.; {Pinto},
  P.A.; {Pizagno}, J.L.; {Popescu}, B.; {Prsa}, A.; {Radcka}, V.; {Raddick},
  M.J.; {Rasmussen}, A.; {Rau}, A.; {Rho}, J.; {Rhoads}, J.E.; {Richards},
  G.T.; {Ridgway}, S.T.; {Robertson}, B.E.; {Roskar}, R.; {Saha}, A.;
  {Sarajedini}, A.; {Scannapieco}, E.; {Schalk}, T.; {Schindler}, R.;
  {Schmidt}, S.; {Schmidt}, S.; {Schneider}, D.P.; {Schumacher}, G.;
  {Scranton}, R.; {Sebag}, J.; {Seppala}, L.G.; {Shemmer}, O.; {Simon}, J.D.;
  {Sivertz}, M.; {Smith}, H.A.; {Allyn Smith}, J.; {Smith}, N.; {Spitz}, A.H.;
  {Stanford}, A.; {Stassun}, K.G.; {Strader}, J.; {Strauss}, M.A.; {Stubbs},
  C.W.; {Sweeney}, D.W.; {Szalay}, A.; {Szkody}, P.; {Takada}, M.; {Thorman},
  P.; {Trilling}, D.E.; {Trimble}, V.; {Tyson}, A.; {Van Berg}, R.; {Vand en
  Berk}, D.; {VanderPlas}, J.; {Verde}, L.; {Vrsnak}, B.; {Walkowicz}, L.M.;
  {Wand elt}, B.D.; {Wang}, S.; {Wang}, Y.; {Warner}, M.; {Wechsler}, R.H.;
  {West}, A.A.; {Wiecha}, O.; {Williams}, B.F.; {Willman}, B.; {Wittman}, D.;
  {Wolff}, S.C.; {Wood-Vasey}, W.M.; {Wozniak}, P.; {Young}, P.; {Zentner}, A.;
  {Zhan}, H.
\newblock {LSST Science Book, Version 2.0}.
\newblock {\em arXiv e-prints} {\bf 2009}, p. arXiv:0912.0201,
  \href{http://xxx.lanl.gov/abs/0912.0201}{{\normalfont
  [arXiv:astro-ph.IM/0912.0201]}}.

\bibitem[Ando \em{et~al.}(2019)Ando et~al.]{Ando:2019rvr}
Ando, S.; others.
\newblock {Discovery prospects of dwarf spheroidal galaxies for indirect dark
  matter searches} {\bf 2019}.
\newblock  \href{http://xxx.lanl.gov/abs/1905.07128}{{\normalfont
  [arXiv:astro-ph.CO/1905.07128]}}.

\bibitem[Moiseev \em{et~al.}(2015)Moiseev et~al.]{Moiseev:2015lva}
Moiseev, A.A.; others.
\newblock {Compton-Pair Production Space Telescope (ComPair) for MeV Gamma-ray
  Astronomy} {\bf 2015}.
\newblock  \href{http://xxx.lanl.gov/abs/1508.07349}{{\normalfont
  [arXiv:astro-ph.IM/1508.07349]}}.

\bibitem[De~Angelis \em{et~al.}(2016)De~Angelis et~al.]{DeAngelis:2016slk}
De~Angelis, A.; others.
\newblock {The e-ASTROGAM mission (exploring the extreme Universe in the
  MeV-GeV range)} {\bf 2016}.
\newblock  \href{http://xxx.lanl.gov/abs/1611.02232}{{\normalfont
  [arXiv:astro-ph.HE/1611.02232]}}.

\bibitem[Tavani \em{et~al.}(2018)Tavani et~al.]{DeAngelis:2017gra}
Tavani, M.; others.
\newblock {Science with e-ASTROGAM: A space mission for MeV–GeV gamma-ray
  astrophysics}.
\newblock {\em JHEAp} {\bf 2018}, {\em 19},~1--106,
  \href{http://xxx.lanl.gov/abs/1711.01265}{{\normalfont
  [arXiv:astro-ph.HE/1711.01265]}}.
\newblock
  doi:{\changeurlcolor{black}\href{https://doi.org/10.1016/j.jheap.2018.07.001}{\detokenize{10.1016/j.jheap.2018.07.001}}}.

\end{thebibliography}

\end{document}